\documentstyle[12pt,psfig]{article}

\newcommand{\Sl}[1]{{}}                              

\renewcommand{\theequation}{\thesection.\arabic{equation}}

\newcommand{\beq}[1]{\Sl{#1}\begin{equation}\if#1\empty\else\label{#1}\fi}
\newcommand{\eeq}{\end{equation}}
\newcommand{\beqa}[1]{\Sl{#1}\begin{eqnarray}\if#1\empty\else\label{#1}\fi}
\newcommand{\eeqa}{\end{eqnarray}}

\newcommand{\bm}[1]{\mbox{\boldmath$ #1 $}}
\newcommand{\bmc}{{\bm c}}

\newcommand{\bmq}{{\bm q}}
\newcommand{\bmr}{{\bm r}}
\newcommand{\bmP}{{\bm P}}
\newcommand{\bmzero}{{\bm 0}}

\newcommand{\hatA}{{\widehat A}}
\newcommand{\hatG}{{\widehat G}}
\newcommand{\nbar}{\overline{n}}
\newcommand{\llangle}{\langle\!\langle}
\newcommand{\rrangle}{\rangle\!\rangle}

\title{Generalized Boltzmann Equation for Lattice Gas Automata\footnote{
preprint THU-94/10, comp-gas/9404002}
}
\author{
H. J. Bussemaker and M. H. Ernst\\
Institute for theoretical physics\\
University of Utrecht, The Netherlands\\[2mm]
and\\[2mm]
J. W. Dufty\\
Physics Department, University of Florida\\
Gainesville, Florida, U.S.A.
}
\date{\small submitted to J.\ Stat.\ Phys., April 1994}

\begin{document}

\maketitle
\vspace{5mm}
\begin{abstract}
\noindent
In this paper, for the first time a theory is formulated that predicts
velo\-city and spatial correlations between occupation numbers that occur in
lattice gas automata violating semi-detailed balance.
Starting from a coupled BBGKY hierarchy for the $n$-particle distribution
functions, cluster expansion techniques are used to derive approximate
kinetic equations.
In zeroth approximation the standard nonlinear Boltzmann equation is obtained;
the next approximation yields the ring kinetic equation, similar to that for
hard sphere systems, describing the time evolution of pair correlations.
As a quantitative test we calculate equal time correlation functions in
equilibrium for two models that violate semi-detailed balance.
One is a model of interacting random walkers on a line, the other one is a
two-dimensional fluid type model on a triangular lattice.
The numerical predictions agree very well with computer simulations.\\[4mm]
\end{abstract}

\newpage

\setcounter{equation}{0}
\setcounter{section}{0}
\section{Introduction}

In the theory of lattice gas automata (LGA's) the Gibbs distribution is
known to be the unique equilibrium distribution if the dynamics satisfy
the so-called semi-detailed balance \cite{G6} or Stueckelberg
\cite{Stueckelberg} condition. The Gibbs distribution depends on the
phase space variables {\em only} through globally conserved quantities,
such as the total number of particle $N$, the total momentum $\bmP$ or
for thermal models the total energy $H$.
If the collision rules are strictly local, involving particles on the same
node only, then the Gibbs equilibrium state is completely factorized, i.e.\
the occupations of velocity channels at the same or at different nodes are
completely uncorrelated.

However, in LGA's that violate the condition of semi-detailed balance
equal time correlations are know to exists in equilibrium between occupation
numbers of different velocity channels on the same or on different nodes
\cite{Dubrulle,Bussemaker92,Coevorden,Henon94}.
The existence of such equilibrium correlations prevents the equilibrium
distribution from being the Gibbs distribution, and therefore standard
equilibrium statistical mechanics does not apply to LGA's that violate
(semi)-detailed balance.
In the existing literature a quantitative or qualitative understanding
of the structure of the equilibrium distribution for such models is
entirely lacking.
Fundamental questions are: is there a unique equilibrium state, or can there
be several stationary states? To what extent does the final state depend
on the initial state and on the details of the collision rules?
How does the system approach the correlated equilibrium state.
How can one calculate transport properties and time correlation functions?

The aim of the present paper is to provide a theory for calculating time
dependent distribution and correlation functions
in LGA's that violate semi-detailed balance.
Using the general ideas developed in the classical
theories of fluids, we construct a BBGKY hierarchy for the distribution
functions, and obtain by means of cluster expansion techniques a generalized
Boltzmann equation for the single particle distribution function,
coupled to the so-called ring equation for the pair correlation function.
We are specially interested in calculating the structure of a spatially
uniform equilibrium state.

The organization of the paper is as follows. In section~2 we derive the first
two equations of a coupled hierarchy for the joint distribution functions,
and we introduce a cluster expansion around mean field theory that leads to
a closure of the hierarchy at the level of two-point correlation functions.
In section~3 the coupled kinetic equations are specialized to spatially
uniform stationary states, and equations are constructed for the stationary
values of the single particle distribution function and the pair correlation
function; an iterative scheme for evaluating these equations numerically
is discussed in section~4.
To establish the accuracy of the theory, we apply in section~5 the theory
to a model of interacting random walkers on a line, and to a fluid-type
model on the triangular lattice, and compare the theoretical predictions
with detailed computer simulations performed for both models.
We end with some conclusions and remarks about future developments in
section~6.

\setcounter{equation}{0}
\setcounter{section}{1}
\section{Kinetic theory}
\subsection{Microdynamic equation}
We consider a lattice gas automaton defined on a regular $d$-dimen\-sional
lattice ${\cal L}$ with periodic boundary conditions, containing $V=L^d$ nodes.
On each node $\bmr\in\cal L$ there exist $b$ allowed velocity channels
$\bmc_1,\bmc_2,\ldots,\bmc_b$, corresponding to the nearest neighbor lattice
vectors; there may also be rest particles with $|\bmc_i|=0$.
The system evolves at discrete time steps $t=0,1,\ldots$.
The microscopic configuration of the system at time $t$ is given in terms of
occupation numbers $n_i(\bmr,t)=\{0,1\}$, denoting the absence or presence of
a particle in velocity channel $(\bmr,\bmc_i)$.
The state of node $\bmr$ is denoted by $n(\bmr,t)=\{n_i(\bmr,t); i=1,2,..,b\}$.

One time step in the evolution involves an instantaneous
{\em collision} step which transforms a precollision state $n(\bmr,t)$ into a
postcollision state $n^*(\bmr,t)$ on all nodes $\bmr$ independently.
It is followed by a {\em propagation} step,
\beq{a4a}
  n_i(\bmr+\bmc_i,t+1) = n^*_i(\bmr,t),
\eeq
during which a particle at node $\bmr$ is moved to the nearest neighbor node
$\bmr+\bmc_i$ in the direction of its velocity $\bmc_i$.
The collision step is specified in terms of a matrix of {\em transition
probabilities} $A_{s\sigma}$ from a precollision or in-state
$s=\{s_1,s_2,\ldots,s_b\}$ to a postcollision or out-state
$\sigma=\{\sigma_1,\sigma_2,\ldots,\sigma_b\}$.
The matrix $A_{s\sigma}$ satisfies the {\em normalization} condition
\beq{a1}
  \sum_\sigma A_{s\sigma} = 1.
\eeq
There are additional constraints imposed on the transition matrix
$A_{s \sigma}$ due to the presence of local conservation laws,
which are most easily expressed in terms of the
{\em collisional invariants} $a_i=\{1,\bmc_i,\frac{1}{2}c_i^2,...\}$,
as
\beq{a1a}
  \sum_i a_i n_i(\bmr,t) = \sum_i a_i n^*_i(\bmr,t).
\eeq
The local conservation laws imply global conservation of
the total quantities
${\cal H}=\{N,\bmP,H,\ldots\} \equiv  \sum_{r,i} a_i n_i(\bmr,t)$.
In diffusive LGA's only the total number of particles, $N$, is conserved;
in fluid-type LGA's the total momentum, $\bmP$, is conserved in addition.
There also exist thermal models where the total energy $H$ is conserved.
Furthermore, the majority of LGA's have additional (spurious) global invariants
\cite{staggered}, which are mostly `staggered' in space and time. They
can be included in the present discussion in a similar manner.

Much is known about the properties of LGA's that satisfy the condition of
{\em semi-detailed balance}
\beq{a2}
  \sum_s A_{s\sigma} = 1,
\eeq
also known as the Stueckel\-berg condition \cite{Stueckelberg},
or the stronger {\em detailed balance} condition,
\beq{a3}
   A_{s\sigma} = A_{\sigma s}.
\eeq
which implies the semi-detailed balance condition (\ref{a2}) on account of
the normalization condition (\ref{a1}).
If condition (\ref{a2}) is satisfied then the equilibrium distribution for
the model is the Gibbs distribution, which only depends on the globally
conserved quantities $\cal H$.\footnote{
The semi-detailed balance and detailed balance conditions with respect to an
arbitrary phase space distribution function $P_0(s)$ have respectively
the forms, $\sum_s P_0(s) A_{s \sigma} =
P_0(\sigma)$ and $ P_0(s) A_{s \sigma} = A_{\sigma s} P_0(\sigma)$.
The phase space distribution in the Gibbs state depends on the variables $
s(\bmr)$ only through locally conserved quantities. Consequently
$P_0(s) = P_0(\sigma)$ and the (semi)-detailed balance conditions reduce to
the equations in the text.
}

A comment on the difference between semi-detailed balance and detailed
balance is appropriate.
We group all states $s(\bmr)$ that can be mapped into one another by lattice
symmetries (rotations, reflections) into the same
{\em equivalence class} \cite{Bussemaker92}.
Then, suppose that for any given in-state, $s(\bmr)$, transitions are allowed
to out-states, $\sigma(\bmr)$, belonging to at most $M$ different equivalence
classes.
As long as $M \leq 2$ there is no distinction between semi-detailed balance
and detailed balance.
Almost all LGA's used in the literature have $M \leq 2$. Notable exceptions
are the FCHC models \cite{Henon92}.

In what follows, no condition apart from normalization will be imposed on
the transition matrix $A_{s\sigma}$. Of course, all the results that will
be obtained in the rest of this paper will hold for LGA's that satisfy
semi-detailed balance as well, being just a special case of the more general
class of LGA's violating semi-detailed balance.

\subsection{Hierarchy equations}

To construct the equations of motion for the occupation numbers we observe that
the occupation numbers {$n(\bmr,t)$} and {$n^*(\bmr, t)$} are related through
the transition matrix $A_{s \sigma}$ as,
\beq{a4}
 n^*_i(\bmr,t) = \sum_{\sigma s}
 \sigma_i \hatA_{s \sigma} \delta(s,n(\bmr,t)),
\eeq
where $\delta(s,n)=\prod_j \delta(s_j,n_j)$ is a product of $b$ Kronecker
delta functions.
For the product of two occupation numbers on the same node we have
\beq{a4b}
 n^*_i(\bmr,t)  n^*_j(\bmr,t)
  	= \sum_{\sigma s} \sigma_i \sigma_j \hatA_{s \sigma}
	\delta(s,n(\bmr,t)).
\eeq
For every node $\bmr$ and every time $t$ the Boolean realization
$\hatA_{s \sigma}=\hatA_{s \sigma}(\bmr,t)=\{0,1\}$ of the
$(s \rightarrow \sigma)$--transition
is drawn from an $(\bmr,t)$--independent probability distribution with
expectation value $\llangle \hatA_{s \sigma} \rrangle = A_{s \sigma}$.

The hierarchy equations describe the time dependence of the distribution
functions, i.e. the
expectation value of products of occupation numbers. The single particle and
two particle distribution functions,
\beqa{a5a}
f_i(\bmr,t) &=& \langle n_i(\bmr,t) \rangle \nonumber \\
f^{(2)}_{ij}(\bmr,\bmr',t) &=& \langle n_i(\bmr,t) n_j(\bmr',t)\rangle,
\eeqa
are defined as averages $\langle\ldots\rangle$ over an arbitrary distribution
of initial occupation numbers, $\{n(\bmr,0)\}$.
Note that $f^{(2)}_{ii}(\bmr,\bmr,t) = f_i(\bmr,t)$.

We now introduce the single node and two node distribution functions,
\beqa{a10}
  p(s,\bmr,t) &=& \langle \delta(s,n(\bmr,t)) \rangle \\
  \label{a10a}
  p^{(2)}(s,\bmr,s',\bmr',t) &=& \langle \delta(s,n(\bmr,t))
  \delta(s',n(\bmr',t)) \rangle,
\eeqa
By taking the average of (\ref{a4}) over an arbitrary distribution of initial
states $\{n_i(\bmr,0)\}$, as well as over the $\hatA$-distribution of
realizations of the transition matrix, and using (\ref{a4a}), we obtain
\beq{a6a}
  f_i(\bmr+\bmc_i,t+1) = \langle n^*_i(\bmr,t) \rangle
  = \sum_{\sigma s} \sigma_i A_{s \sigma} p(s,\bmr,t).
\eeq
To obtain an analogous equation for the pair distribution function we must
distinguish between the cases $\bmr=\bmr'$ and $\bmr\neq\bmr'$.
When $\bmr\neq\bmr'$ the collisions for channels $(\bmr,\bmc_i)$ and
$(\bmr',\bmc_j)$ are independent, and given by (\ref{a4}).
When $\bmr=\bmr'$ however we must use (\ref{a4b}).
The combined result can be written as
\beqa{a8a}
 f^{(2)}_{ij}(\bmr+\bmc_i, \bmr + \bmc_j,t+1)
 &=& \langle n^*_i(\bmr,t) n^*_j(\bmr',t) \rangle = \nonumber\\
	&& \hspace{-50mm} [1-\delta(\bmr,\bmr')] \left\{
	\sum_{\sigma s, \sigma' s'} \sigma_i {\sigma}'_j
	A_{s \sigma} A_{s' \sigma'} p^{(2)}(s,\bmr,s',\bmr',t)  \right\} \\
	&& \hspace{-50mm} \mbox{} + \delta(\bmr,\bmr') \left\{
 	\sum_{\sigma s} \sigma_i \sigma_j A_{s \sigma} p(s,\bmr,t)
	\right\}. \nonumber
\eeqa
The above equations (\ref{a6a}) and (\ref{a8a})
constitute respectively the first two equations of an open
hierarchy of coupled equations for the distribution functions, similar to
the BBGKY hierarchy for continuous systems.
It is straightforward to extend this procedure to construct the hierarchy
equations for $f^{(\ell)}$ with $\ell \geq 3$, although the complexity
increases rapidly.
In this paper the first two equations of the BBGKY hierarchy suffice for our
purpose.

The node distribution functions are related to the $\ell$--particle
distribution
functions in a simple manner. This follows from the following identity, valid
for binary variables,
\beq{a6}
  \delta(s,n) \equiv \prod_j \delta(s_j,n_j)
	= \prod_j n_j^{s_j} (1-n_j)^{1-s_j}.
\eeq
As $s_j$ takes only values 0 or 1, the right hand side is a sum of products of
at most $b$ occupation numbers. The average of (\ref{a6}), $p(s,\bmr,t)$, is
therefore a linear combination of $\ell$-particle distribution functions
$f^{(\ell)}$ with $\ell=1,2,..,b$, all referring to the same node, and with
coefficients that are either $+1$ or $-1$. Similarly $p^{(2)}$ is a linear
combination of distribution functions.

To understand the differences between the BBGKY hierarchy for discrete LGA's
and continuous systems we note the following. In continuous systems usually
only two-body interactions (additive forces) are considered; consequently
the RHS of the $\ell$-th hierarchy equation contains only $(\ell+1)$-particle
distribution functions. If one would include $s$-body interactions
($s=2,3,4,..,b$) the $\ell$-th hierarchy equation would contain
$(\ell+b-1)$-particle distribution functions.
In the LGA case $b$-body interactions simultaneously occur on each node $\bmr$,
where $b$ is the number of velocity channels at a node.
Therefore the first hierarchy equation already involves up to $b$-particle
distribution functions; the second equation even involves up to $2b$-particle
functions.

\subsection{Cluster expansion}
An approximate closure of the hierarchy equations can be obtained by
making a cluster expansion in terms of two-, three-, etc.\ point
correlation functions and retaining correlation
functions up to a certain order while neglecting the higher order ones.
This procedure leads to approximate kinetic equations.

The cluster functions $G$ are defined through the so-called Ursell expansion,
\beqa{b1b}
f^{(2)}_{ij}(\bmr,\bmr') &=& f_i(\bmr) f_j(\bmr')
	+ G_{ij}(\bmr,\bmr') \nonumber \\
f^{(3)}_{ij \ell}(\bmr,\bmr',\bmr'') &=& f_i(\bmr) f_j(\bmr') f_\ell(\bmr'')
	+ f_i(\bmr) G_{j\ell}(\bmr',\bmr'')
	+ f_j(\bmr') G_{i\ell}(\bmr,\bmr'') \nonumber\\
	&& \mbox{}
	+ f_\ell(\bmr'') G_{ij}(\bmr,\bmr') + G_{ij\ell}(\bmr,\bmr',\bmr''),
\eeqa
etc, where in general all channels $(\bmr,\bmc_i)$, $(\bmr',\bmc_j)$,
and $(\bmr'',\bmc_l)$ are different.
Solution of the recursion relations yields the pair and triplet functions
in terms of the $f^{(\ell)}$.
The cluster functions conveniently can be written in terms of fluctuations,
\beq{a11a}
  \delta n_i(\bmr,t) =  n_i(\bmr,t) -  f_i(\bmr,t).
\eeq
For instance, for the two point function we have
\beq{a11}
G_{ij}(\bmr,\bmr',t) =  f^{(2)}_{ij}(\bmr,\bmr',t)
 - f_i(\bmr,t) f_j(\bmr',t)
	= \langle \delta n_i(\bmr,t) \delta n_j(\bmr',t) \rangle,
\eeq
and similarly we can write
$G_{ij\ell}(\bmr,\bmr',\bmr'',t) = \langle \delta n_i(\bmr,t)
\delta n_j(\bmr',t) \delta n_\ell(\bmr'',t) \rangle$.
It is important to note that the diagonal elements for
$(\bmr,\bmc_i) = (\bmr',\bmc_j)$ are completely determined by $f_i(\bmr,t)$,
due to the Boolean character of $n_i(\bmr,t)$,
\beq{a12}
  G_{ii}(\bmr,\bmr,t)
	= \langle (\delta n_i(\bmr,t))^2 \rangle
	= f_i(\bmr,t) (1-f_i(\bmr,t)) \equiv g_i(\bmr,t).
\eeq

To perform the cluster expansion of the hierarchy equations the cluster
expansion of the node distribution functions $p(s)$ and $p^{(2)}(s,s')$ is
needed.  In appendix~A the expansion up to terms linear in the pair
correlations is worked out detail. It amounts to obtaining approximate
expressions for $p$ and $p^{(2)}$ in terms of $f$ and $G$.
For the single node distribution function the results is
\beq{b2}
  p(s,\bmr,t)	= F(s,\bmr,t) \left\{1+\sum_{k<\ell}
	\frac{\delta s_k(\bmr,t) \delta s_l(\bmr,t)}{g_k(\bmr,t) g_l(\bmr,t)}
	    G_{k\ell}(\bmr,\bmr,t) + \ldots \right\},
\eeq
as shown in appendix~A.
Here $F(s,\bmr,t)$ is the {\em completely factorized} single node
distribution function,
\beq{b3}
  F(s,\bmr,t) = \prod_j (f_j(\bmr,t))^{s_j} (1-f_j(\bmr,t))^{1-s_j}.
\eeq
In a similar way we can expand the two-node distribution function
(see eq.~(\ref{m4}) of appendix~A).
Substitution of these expansions in the hierarchy equations yields coupled
approximate equations of motion for $f_i(\bmr,t)$ and $G_{ij}(\bmr,\bmr',t)$,
given by (\ref{m2}) together with (\ref{m4}), (\ref{m6}), (\ref{m7}) and
(\ref{m10}).

The kinetic equation in {\em zeroth} approximation
is obtained by neglecting $G_{ij}$ in (\ref{b2}),
or equivalently by replacing the
distribution functions $p(s,\bmr,t)$ in (\ref{a6a})
by the fully factorized $F(s,\bmr,t)$.
In that case the first hierarchy equation reduces to the
{\em nonlinear Boltzmann equation},
\beq{b7}
  f_i(\bmr+\bmc_i,t+1) - f_i(\bmr,t) =
 \Omega^{(1,0)}_i(f(\bmr,t)),
\eeq
with the nonlinear collision operator $\Omega^{(1,0)}_i(f)$ defined in
(\ref{m3}).
This is the standard form of the nonlinear lattice Boltzmann approximation
(mean field theory) for lattice gas automata \cite{G6}.
The way in which this equation has been derived closely parallels the
derivation of the nonlinear Boltzmann equation for hard spheres from the
corresponding BBGKY hierarchy \cite{Ernst81}, where one replaces the pair
distribution function $f^{(2)}$ in the first hierarchy equation by a product of
single particle distribution functions, or equivalently, one neglects the two
point correlation function.

\subsection{Generalized Boltzmann and Rings}

In {\em first} approximation the terms linear in $G_{k\ell}$ are included in
the kinetic equations, but higher order terms are neglected.
The nonlinear Boltzmann equation (\ref{b7}) is extended with a term linear
in $G$, yielding the {\em generalized Boltzmann equation},
\beq{b8}
  f_i(\bmr+\bmc_i,t) - f_i(\bmr,t) =
	 \Omega^{(1,0)}_i(f(\bmr,t)) + \sum_{k<\ell}
	\Omega^{(1,2)}_{i,k\ell}(f(\bmr,t)) G_{k\ell}(\bmr,\bmr,t).
\eeq
It describes corrections to the mean field equation (\ref{b7}), caused by the
correlations $G_{ij}$ between the colliding particles. These
correlations can be calculated from the lowest approximation to the second
hierarchy equation. The derivation is given in appendix~A.
The result is the so-called {\em ring kinetic equation},
\beq{b9}
  G_{ij}(\bmr+\bmc_i,\bmr'+\bmc_j,t+1) =
	\omega_{ij,k\ell}(f,f') G_{k\ell}(\bmr,\bmr',t) +
	 \delta(\bmr,\bmr') B_{ij}(\bmr,t),
\eeq
where $\omega$ is the two particle collision operator, given by
\beq{b10b}
	\omega_{ij,k\ell}(f,f') = \{1 + \Omega^{(1,1)}(f(\bmr,t))\}_{ik}
	    \{1 + \Omega^{(1,1)}(f(\bmr',t))\}_{j\ell},
\eeq
with $\Omega^{(1,1)}_i(f(\bmr,t))$ defined in (\ref{m3}).
The on-node source term $B_{ij}(\bmr,t)$ is defined by (\ref{m6})
and (\ref{m10}) and depends on both $f_i(\bmr,t)$ and $G_{ij}(\bmr,\bmr,t)$.

The ring equation (\ref{b9}) in combination with the generalized Boltzmann
equation (\ref{b8}) form a closed set of equations for the
functions $\{f_i(\bmr,t)$, $G_{ij}(,\bmr,\bmr', t)\}$.
These approximate kinetic equations
(\ref{b8}) and (\ref{b9}) still obey the global (standard and spurious)
conservation laws, since the {\em orthogonality} condition with respect to
the collisional invariants $a_i$,
\beq{b13}
  \sum_{ij} a_i a_j B_{ij} = 0,
\eeq
is satisfied for $B_{ij}$, as defined by (\ref{m6}) and (\ref{m10}).
We finally note that $\Omega^{(1,1)}_{ij}(f)$ is the linearized version
of the Boltzmann collision operator, obtained by Taylor expansion,
in powers of $\delta\! f_i$,
of the coefficient $\Omega^{(1,0)}(f) $ in (\ref{m3})
around some arbitrary distribution $f_i=f_i(\bmr,t)$,
\beq{b14}
 \Omega^{(1,0)}_i (f+\delta\! f) = \Omega^{(1,0)}_i(f) + \sum_j
 \Omega^{(1,1)}_{ij}(f) \delta\! f_j + \ldots,
\eeq

We discuss the assumptions and the relevance of the generalized Boltzmann
and ring kinetic equation for LGA's that satisfy or violate semi-detailed
balance.
The generalized Boltzmann equation and ring equation also describe
non-equilibrium phenomena, such as transport coefficients. Application of the
Chapman-Enskog method to the system of equations (\ref{b8}) and  (\ref{b9})
will yield its normal solution and the transport coefficients. For the special
case of detailed balance models the corrections to the Boltzmann transport
coefficients resulting from our equations have already been calculated
\cite{Velzen93}, and the agreement with simulation results is quite good
\cite{Gerits}. For LGA's violating detailed balance the numerical calculation
of transport coefficients beyond the Boltzmann approximation has become
feasible with the present theory, but a detailed analysis has yet to be
performed.

Moreover, the present ring kinetic theory describes the algebraic long time
tails of the velocity correlation and other current correlation functions,
whereas the Boltzmann equation predicts only exponential decay. Also here the
analytic predictions \cite{Kirkpatrick91} are in excellent agreement with the
result of simulated long time tails for LGA's satisfying detailed balance
\cite{Frenkel89}.
No long time tail measurements nor theoretical calculations have been
reported in the literature for LGA's violating semi-detailed balance.

The cluster expansions of the first two hierarchy equations also
generate terms of degree $b$ in the pair functions. In principle these terms
might be included in (\ref{b8}) and (\ref{b9}), but a numerical solution does
not seem feasible at present. Moreover, such extensions seem unnecessary, to
judge from the good agreement between theory and simulation results reported
above \cite{Gerits} and in the following sections.

The equation of motion (\ref{b9}) for the equal time correlation function
$G_{ij}(\bmr,\bmr',t)$ can also be derived from the
Boltzmann-Langevin equation, obtained by adding a stochastic source
(noise) to the nonlinear Boltzmann equation. In the weak noise
limit the equation (\ref{b9}) for $G_{ij}(\bmr,\bmr',t)$
follows, where the source term $B_{ij}$ is
the covariance of the stochastic noise source \cite{Dufty94,Ernst81}.
The form of this source is not specified in such a phenomenological
approach without further input, whereas the derivation here provides
the explicit representation  of $B_{ij}(\bmr,t)$
as a function of $f_{i}(\bmr,t)$ and $G_{ij}(\bmr,\bmr,t)$.

The equations above were derived under the assumption that triplet- and
higher correlations, as well as products of pair correlations are small.
The justification of these assumptions can only be given a posteriori
by comparing the theoretical predictions with results from computer
simulations. This will be done in the remainder of this paper for the special
case of equilibrium correlations.

\setcounter{equation}{0}
\setcounter{section}{2}
\section{Equilibrium correlations}
The purpose of the present
section is to study the approximate equations for the
equilibrium correlation functions, and obtain solutions under the assumption
that a {\em spatially uniform equilibrium} state exists. The single particle
distribution function will be denoted by $f_i(\bmr,\infty)=f_i$ and the
pair correlations by $G_{ij}(\bmr,\bmr',\infty)=G_{ij}(\bmr-\bmr')$.
As the on-node correlations play a special role, it is convenient to denote
them simply by $G_{ij}=G_{ij}(\bmr,\bmr,\infty)$.

For LGA's satisfying semi-detailed balance it is well known that the
equilibrium distributions are
completely factorized \cite{G6,Bussemaker92}.  Therefore the solutions should
reduce to $G_{ij}(\bmr-\bmr') = \delta(\bmr,\bmr') \delta_{ij} g_i$.  This is
indeed the case, as will be briefly explained at the end of this section.
In LGA's violating semi-detailed balance, Eq.~(\ref{b9}) gives a quantitatively
correct prediction for the on- and off-node correlations, extensively measured
in the literature \cite{Dubrulle,Bussemaker92,Coevorden,Henon94}.
This is the main subject of this and the next sections.

The kinetic equation (\ref{b8}) for $f_i$ simplifies to
\beq{c1}
\Omega^{(1,0)}_i + \sum_{k<\ell} \Omega^{(1,2)}_{i,k\ell}
	G_{k\ell} = 0.
\eeq
The ring equation (\ref{b9}) for the pair correlations can also be simplified.
After introduction of the Fourier transform,
\beq{c2}
  \hatG_{ij}(\bmq) = \sum_\bmr e^{-{\textstyle i}\bmq\cdot\bmr} G_{ij}(\bmr),
\eeq
it takes the form
\beq{c3}
  \hatG_{ij}(\bmq) - s_{ij}(\bmq) \omega_{ij,k\ell} \hatG_{k\ell}(\bmq) =
	s_{ij}(\bmq) B_{ij}
\eeq
or in matrix notation
\beq{c4}
  \{1 - s(\bmq) \omega \} \hatG(\bmq) = s(\bmq) B,
\eeq
where $\hatG(\bmq)$, $G$, and $B$ are $b^2$-dimensional vectors with
components $\hatG_{ij}(\bmq)$, etc., and $\omega$, $s(\bmq)$, and $1$
are $b^2 \times b^2$ matrices with elements $\omega_{ij,k\ell}$,
$s_{ij,k\ell}(\bmq)=s_{ij}(\bmq) \delta_{ik} \delta_{j\ell}$, and
$1_{ij,k\ell} = \delta_{ik} \delta_{j\ell}$.
The   pair streaming operator is defined as
\beq{c5}
  s_{ij}(\bmq) = \exp[-i \bmq\cdot(\bmc_i-\bmc_j)] .
\eeq
In the spatially uniform equilibrium state
the quantity $B_{ij}$ simplifies to,
\beq{c6c}
  B_{ij} = \Omega_{ij}^{(2,0)} +
	\sum_{k < \ell}\Omega^{(2,2)}_{ij,k\ell} G_{k\ell}
 	+  \sum_{k \ell}(1-\omega)_{ij,k\ell} G _{k\ell}.
\eeq
Note that due to (\ref{c1}) the contribution $B^{(3)}_{ij}$ in (\ref{m6})
vanishes identically.

In appendix B we discuss the technicalities of obtaining a well-defined closed
equation (\ref{c19}) for the {\em on-node} pair correlation function $G_{ij}$
or more conveniently for the {\em excess} correlation function
$C_{ij} = G_{ij} -\delta_{ij} g_i$.
This equation reads
\beq{c9c}
  C = R \{ \Omega^{(2,2)} + (1-\omega) \} C + R \Omega^{(2,0)} + J,
\eeq
where the ring operator $R$ and the term $J$ are given by (\ref{c15}) and
(\ref{c20}) respectively.
Most technical difficulties are related with finite size effects that give
deviations for small system sizes $V=L^d$.
In the thermodynamic limit ($V \rightarrow\infty$) many complications
disappear:
the term $J$ vanishes, and the ring operator becomes
\beq{c16c}
  R = \lim_{V\rightarrow\infty} \frac{1}{V} \sum_{\bmq \in \rm 1BZ}
  	\frac{1}{1-s(\bmq)\omega} s(\bmq)
    = v_0 \int_{\rm 1BZ} \frac{d\bmq}{(2\pi)^d}
	\frac{1}{1-s(\bmq)\omega} s(\bmq),
\eeq
where summation and integration are restricted to the first Brillouin zone
(1BZ) of the reciprocal lattice ${\cal L}^*$ and where $v_0$ denotes the
volume of a unit cell on the lattice ${\cal L}$
(e.g.\ $v_o=1$ for the square and $v_0=\frac{1}{2}\sqrt{3}$ for the
triangular lattice).

Once the on-node correlations are known from the solution (\ref{c9c}), then
$\hatG_{ij}(\bmq)$ can be calculated from (\ref{c4}),
and the full $\bmr$-dependent correlation function
$G_{ij}(\bmr)$ follows by inverse Fourier transformation of $\hatG_{ij}(\bmq)$,
\beq{c16}
	G_{ij}(\bmr) = \frac{1}{V} \sum_\bmq
	e^{{\textstyle i}\bmq\cdot\bmr} \hatG_{ij}(\bmq).
\eeq
In equilibrium the pre- and postcollision correlations are related by
\beq{c17}
	G^*_{ij}(\bmr) = G_{ij}(\bmr+\bmc_i-\bmc_j),
\eeq
because of the relation $  n_i(\bmr+\bmc_i,t+1) = n^*_i(\bmr,t)$
between pre- and postcollision occupation numbers.

Before concluding this section we make some further comments on the ring
equation (\ref{c9c}).  The term $R\Omega^{(2,0)}$ on the right hand side
yields the {\em simple ring} approximation to the (precollision) correlation
function $C$. The quantity $\Omega^{(2,0)}_{ij}(f^0)$, defined in (\ref{m13}),
constitutes the lowest order approximation to the postcollision correlations.
It represents the on-node postcollision correlations, created by a single
collision from a completely factorized precollision state.
This quantity was calculated earlier in Ref.~\cite{Bussemaker92}, and
agreement within 10\% was observed in computer simulations on a triangular
lattice gas violating semi-detailed balance.
The present theory enables one to extend these calculations to
the precollision correlations, which are built up by collective phenomena,
acting on large spatial and temporal scales.
The ring operator (\ref{c15}) propagates these
two particle correlations to other nodes through uncorrelated single particle
motion (Boltzmann propagators), and recollects this information on a single
node.

The term $R\{\Omega^{(2,2)}+(1-\omega)\}C$ on the right hand side of
(\ref{c9c}) represents the repeated ring collisions. Numerical evaluation of
(\ref{c9c}) in the next section will show that the repeated
rings represent only small corrections to the equilibrium correlations,
as calculated from the dominant simple ring approximation, $R \Omega^{(2,2)}$.
Nonvanishing equilibrium velocity and position correlations are only found in
LGA's that violate the condition (\ref{a2}) of semi-detailed balance.
We will return to such LGA's in the next section.

In the remainder of this section we show that the excess correlation functions
$C_{ij}$ are vanishing in LGA's that satisfy the semi-detailed balance
condition.
If mass and possibly momentum
is conserved in a LGA, the equilibrium solution is $f_i=\rho/b$  with $\rho$
the average number of particles per node. This can be verified from the
nonlinear Boltzmann equation, showing $\Omega^{(1,0)}_i(f^0)=0$.

The factorized node distribution (\ref{b3}) obeys the relation,
$F_0(s)=F_0(\sigma)$, where $s$ and $\sigma$ are respectively pre-
and postcollision occupation numbers.
This is so because $F_0(s)$ depends only on the conserved number of
particles, $\sum_i s_i(\bmr)=\sum_i \sigma_i(\bmr)$ on a single node.
The relation $F_0(s)=F_0(\sigma)$ also implies that
$\Omega^{(2,0)}_{ij}(f^0)=0$
in (\ref{m13}) for semi-detailed balance models. This can be verified by
using in the first and second term on the right hand side of (\ref{m13})
respectively the semi-detailed balance condition (\ref{a2}) and the
normalization condition (\ref{a1}).
Consequently, the ring equation (\ref{c19}) gives $C_{ij}=0$ for semi-detailed
balance models. There are no on- or off-node velocity correlations
in the equilibrium state of LGA's that obey the semi-detailed balance
condition, i.e.\
$G_{ij}(\bmr,\bmr') = \delta_{ij} \delta(\bmr,\bmr') f^0_i (1-f^0_i)$,
in full agreement with the Gibbs distribution.

\setcounter{equation}{0}
\setcounter{section}{3}
\section{Numerical evaluation}
So far, we have derived an equation to determine static pair correlations.
This section deals with the numerical evaluation of the theory from a more
operational point of view.  We discuss an iterative scheme for finding the
equilibrium solution numerically, and make some remarks about its uniqueness.

The stationary solution $\{f,G\}$ must satisfy the generalized Boltzmann
equation (\ref{c1}) and  the ring equation (\ref{c9c}). In single speed LGA's
violating semi-detailed balance lattice symmetries require that the single
particle distribution
function is given by $f^0_i= \rho/b$. To calculate the on-node correlation
function from (\ref{c9c}) we first evaluate $\Omega^{(2,0)}$ and
$\Omega^{(2,2)}$ for $f^0_i= \rho/b$, and then solve the linear matrix equation
(\ref{c14}) to obtain the excess correlation function $C_{ij}$.

However, in multi-speed models (for instance, with additional rest particles)
that violate semi-detailed balance the stationary single particle distribution
generally differs from $\rho/b$.
How can one numerically determine the equilibrium solution in the general case?
We have found it efficient to use the following scheme, which amounts to
self-consistent iteration of (\ref{c1}) and (\ref{c9c}):
\begin{enumerate}

\item Specify an accuracy $\epsilon \ll 1$.

\item Find the stationary distribution $f^0_i$ in Boltzmann approximation,
i.e.\ with all pair correlations $G_{ij}$ set equal to zero.
Starting from $f_i^0=\rho/b$ we repeat the iterative step,
$f(n+1) = f(n) + \Omega^{(1,0)}(f(n))$,
until $|\Omega^{(1,0)}_i|<\epsilon$. Note that $\Omega^{(1,0)}_i$ must be
re-calculated at every iteration step.

\item Calculate the (unique) solution $G_{ij}$ of the ring equation
(\ref{c9c}) for given fixed $f_i^0$.

\item Find the stationary solution $f^0_i$ of (\ref{c1})
for given fixed $G_{ij}$ as found in step (3).
This can be done by repeating the iterative step $f(n+1)= f(n) + I(f(n))$,
with $I=\Omega^{(2,0)}+\Omega^{(2,2)} G$, until $|I|<\epsilon$.
Again $I$ must be re-calculated at every iteration step.

\item Repeat steps (3) and (4) until the scheme has converged.

\end{enumerate}
Step (3) in this scheme requires some explanation.  The ring operator $R$ is
calculated using the spectral decomposition (\ref{c6}). This requires a
determination of the $\bmq$-dependent eigenvalues $\lambda_\alpha (\bmq) $ and
the complete bi-orthogonal set of left- and right eigenfunctions $\{
\chi_\alpha
(\bmq), \tilde{\chi}_\alpha (\bmq)\}$ of the two particle propagator
$\gamma(\bmq)=s(\bmq)\omega$. Once $R$ is known, we proceed to
calculate $G_{ij}$ or $C_{ij}$ for given $f^0_i$ by splitting the ring equation
(\ref{c9c}) into an inhomogeneous part that depends only on $f^0_i$, and a part
that depends only on the excess pair correlation functions.
We can write (\ref{c9c}) as
\beq{d1}
   C = M C + K.
\eeq
with $M = R(\Omega^{(2,2)}+1-\omega)$ and $K = R\Omega^{(2,0)}+J$.
Since $C_{ij}=C_{ji}$ is symmetric, the only independent
elements of $C_{ij}$ are those with $i<j$.
These can conveniently be taken together as a vector in a
$\frac{1}{2}b(b-1)$-dimensional space, spanned
by the possible pairs $(ij)$ with $i<j$.
It is therefore appropriate to interpret (\ref{d1}) as a vector
equation, where $C$ and $K$
are $\frac{1}{2} b(b-1)$-dimensional vectors and $M$
is a matrix of equal dimensionality.
Once $K$ and $M$ are known, $C =(1-M)^{-1}K$ can readily be
calculated, provided that $\det(1-M) \neq 0$.
As noted earlier, the term $J$ vanishes in the thermodynamic limit.

At this point we have completed the detailed description of our scheme for
finding the stationary solution to the coupled time evolution equations for
$f_i(t)$ and $G_{ij}(t)$.
Once the equilibrium values of $f^0_i$ and $G_{ij}$ have been found,
$\hatG_{ij}(\bmq)$ and $G(\bmr)$ can be calculated from (\ref{c4}) and
(\ref{c16}).  One can now ask whether the solution that is
obtained with this scheme is unique.  The stationary Boltzmann distribution
$f^0_i$, obtained in step (2), is by definition in the basin of attraction of
the fixed point $\{f^0,G\}$ found using our iterative scheme.
In principle it can not be excluded that there are other stable fixed points
that correspond to physically acceptable equilibrium states as well, since
the $\Omega$-matrices are nonlinear functions of $f=\{f_i\}$.
However, we can systematically
search for other fixed points, since for a given $f$ the corresponding
$G_{ij}(f)$ is uniquely determined by the ring equation.
Then by investigating $I_i(f,G(f))=\Omega^{(1,0)}+\Omega^{(1,2)}G$, as a
function of $f$ we can locate the fixed points, given by the condition
$I_i(f,G(f))=0$.

It is important to stress the difference between two alternative ways of
finding a spatially homogeneous stationary solution to the time evolution
equations, (\ref{b8}) and (\ref{b9}).
The first and most efficient method uses the iterative scheme discussed
in this section, which can be interpreted as a way to find the fixed point
to a mapping $\{f^{(n)},G^{(n)}\}\rightarrow\{f^{(n+1)},G^{(n+1)}\}$.
A second method is to specialize (\ref{b8}) and (\ref{b9}) to the
spatially homogeneous case, which yields another mapping,
$\{f(t),G(\bmr,t)\}\rightarrow\{f(t+1),G(\bmr,t+1)\}$.
If there is more than one stationary solution, then the two schemes will
not necessarily converge to the same fixed point $\{f,G\}$.

\setcounter{equation}{0}
\setcounter{section}{4}
\section{Applications}
\subsection{Interacting random walkers}
The ring kinetic theory for approximate calculation of pair correlations
presented in the previous sections assumes that all higher order correlations
are negligible. It is important to make a quantitative comparison with computer
simulation results to establish the accuracy of the results obtained from a
numerical evaluation of the theory.

\begin{figure}[h]
\centerline{
  \psfig{figure=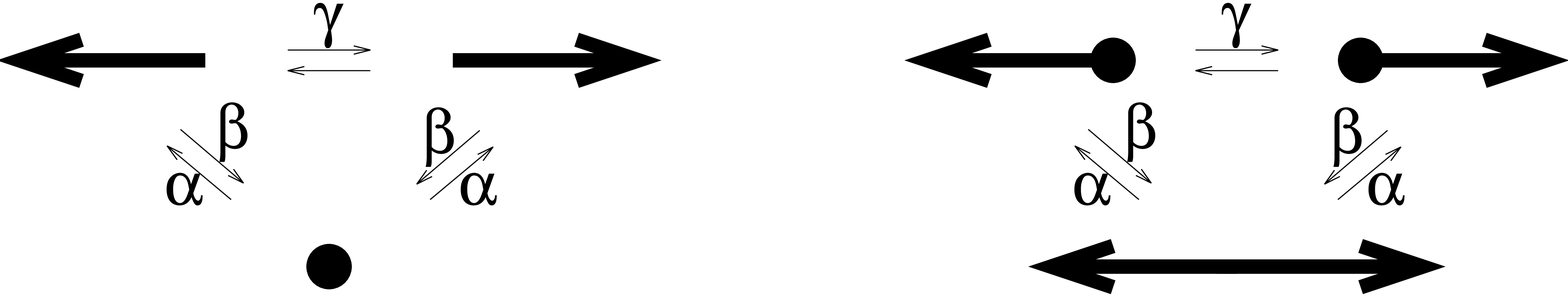,height=2.5cm}
}
\caption{
Transition probabilities in one-dimensional model of interacting random
walkers with mutual exclusion in the same state $\{{\bf r}, {\bf c}_i\}$
with ${\bf c}_i=\{-1,0,1\}=\{\leftarrow,\bullet,\rightarrow\}$.
The left and right diagrams show transitions between states with one and
two particles, respectively.
}
\label{fig1}
\end{figure}

We first consider a model of interacting random walkers on a line. An isolated
random walker executes a persistent random walk, while jumping to nearest
neighbor sites or resting on the same site, i.e. it has three allowed velocity
states, $\bmc_i=\{+1,0,-1\}$ with transition probabilities depending on the
previous jump, as defined in the left diagram of Fig.~\ref{fig1}.

The interaction between the walkers is strictly local, and essentially
determined by the Fermi exclusion rule for different velocity channels, i.e.
$n_i(r,t) = \{0,1\}$. The maximum number of walkers on a site is therefore
three. The transition probabilities from a two-particle in-state $s(r)$ to
an out-state $\sigma(r)$ are defined in the right diagram of Fig.1. There
exists
only a single three-particle state, which remains unchanged under interaction
on
account of the Fermi exclusion rule.

The above model can be described conveniently in terms of a one-dimen\-sional
diffusive LGA, where at every node one- and two-particle transitions
occur with probabilities defined in Fig.1. In principle there are
twelve independent transition probabilities $A_{s \sigma} \neq 0$.
This number is reduced to six if we require that the collision rules be
invariant under reflection ($\bmc_i \rightarrow -\bmc_i$).
One may further impose {\em self-duality}, i.e.\ invariance
under exchange of particles and holes ($n_i \rightarrow \bar{n} \equiv 1-n_i$).
This would
reduce the number of independent transition probabilities to three, so that
$\alpha = \alpha', \beta = \beta', \gamma =\gamma'$.
In the present model the total number of particles is conserved, but total
momentum is not, and there are no staggered or other spurious invariants.
Consequently the model has only a single slow (diffusive) mode.

To illustrate some of the analytical results of the previous sections, and to
discuss the conditions of semi-detailed balance, it is of interest to write
out the microdynamic equation (\ref{a4})
for the occupation numbers $n_+ (r,t)$, $n_0 (r,t)$, and $n_- (r,t)$, referring
respectively to the velocity channels $\{ c_i = +1,0,-1\}$. Let
 $n^*_i(r,t) = n_i(r+c_i,t+1)$ be  the postcollision occupation
number, then
\beqa{f1}
  n^*_+ &=& n_+
	+ (\hat\alpha\; n_0 \nbar_+ - \hat\beta\; n_+ \nbar_0) \nbar_-
	+ \hat\gamma\; (n_-\nbar_+ - n_+ \nbar_-) \nbar_0
	+ \nonumber \\
 	&& \hspace{4mm}
	+ (\hat\beta'\; n_0 \nbar_+ - \hat\alpha'\; n_+ \nbar_0) n_-
	+ \hat\gamma'\; (n_-\nbar_+ - n_+\nbar_-) n_0
	\nonumber \\
n^*_0  &=& n_0
	+ \hat\beta\; (n_+ \nbar_- - n_- \nbar_+) \nbar_0
	- \hat\alpha_r\; n_0 \nbar_+ \nbar_ -
	- \hat\alpha_l\; n_0 \nbar_+ \nbar_ -
	\nonumber \\
 	&& \hspace{4mm}
	+ \hat \alpha^\prime_r \,n_+ n_-\nbar_0
	+ \hat \alpha^\prime_l \,n_+ n_-\nbar_0
	- \hat\beta'\; (n_+ \nbar_- + n_- \nbar_+) n_0,
\eeqa
and a similar relation for $n_-^*$.
The set $\hatA_{s\sigma}=\{\hat{\alpha}_r, \hat{\alpha}_l,
\hat{\alpha}^\prime_r, \hat{\alpha}^\prime_l,
\hat{\beta}, \hat{\beta}^\prime, \hat{\gamma}, \hat{\gamma}^\prime\}$
represents the Boolean realizations, $\hatA_{s \sigma}(r,t)$, discussed
below (\ref{a4}), with expectation values
$A_{s\sigma}=\{\alpha_r, \alpha_l , \alpha^\prime_r, \alpha^\prime_l,
\beta, \beta^\prime, \gamma, \gamma^\prime \}$.
The subscripts $\{r,l\}$ refer to transitions to
states with a particle moving to the right and to the left, respectively.

Consider first the mean field or Boltzmann approximation.
According to (\ref{b7}) the nonlinear Boltzmann equation for $f_i(r,t)$ is
obtained from (\ref{f1}) by replacing $n_i (r,t)$ by $f_i(r,t)$ and the
Boolean variables $\hat{\alpha}$, etc by their expectation values $\alpha$,
etc.. In equilibrium the distribution function $f^0_i$ is characterized by the
average occupation numbers $f_0$ and $f_+ = f_-$ with average density $\rho =
f_0 + 2 f_+$. At a given density $\rho$ the nonlinear Boltzmann equation for
the stationary state reduces to a quadratic equation for $f_0$.

Inspection of (\ref{f1}) shows that the semi-detailed balance condition
(\ref{a3}) is only satisfied if
\beq{f2}
\alpha \alpha^\prime = \beta \beta^\prime,
\eeq
in which case the stationary distribution is given by $z_0/z_+ = \beta/\alpha =
\alpha^\prime / \beta^\prime$, with $z_i = f_i/(1-f_i)$.
We recall that the stationary
distribution is a Gibbs distribution if the phase space density depends only
on the conserved quantities, so that $f^0_i = f_0 = f_+ = f_- =
\frac{1}{3}\rho$
is independent of the velocity channel. Therefore detailed balance with
respect to the Gibbs distribution imposes the condition,
\beq{f3}
\alpha = \beta, \qquad \alpha^\prime = \beta^\prime.
\eeq
With the small set of allowed states $s(r)$ the condition for semi-detailed
balance reduces to that for detailed balance \cite{Bussemaker92}.

If the above detailed balance conditions are violated, there exist
correlations between the occupation numbers of the different channels
$(\bmr,\bmc_i)$, even in the equilibrium state.
The critical quantity, that determines whether equilibrium correlations are
non-vanishing, is matrix element $\Omega^{(2,0)}_{ij} (f^0)$, defined in
(\ref{m13}). It gives for the present model,
\beqa{f4}
\Omega^{(2,0)}_{+-}(f^0) &=& 2(\beta^\prime f_0\overline{f}_+
- \alpha^\prime f_+\overline{f}_+) f_+
\nonumber \\
\Omega^{(2,0)}_{+0} (f^0) &=& \Omega^{(2,0)}_{-0} (f^0) =
(\alpha^\prime \, f_+\overline{f}_0 - \beta^\prime f_0 \overline{f}_+) f_+.
\eeqa
If the transition probabilities satisfy the detailed balance conditions,
then $\Omega^{(2,0)}_{ij}(f^0)$ is vanishing for $\forall i,j$. All excess
on- and off-node correlations vanish, and the stationary distribution is
simply $ f_0 = f_+ = f_- = \frac{1}{3}\rho $.
In case the detailed balance conditions are violated, the elements of
$ \Omega^{(2,0)}(f^0)$ differ from zero, and there exist non-vanishing on- and
off-node correlations in the equilibrium state.

The method developed in the present paper enables one to calculate the
equilibrium distributions $\{ f_i, G_{ij} \}$ numerically, from which the
off-node correlations $G_{ij} (r)$ can be constructed with the help of
(\ref{c16}).
The postcollision correlations $G^*_{ij}(r)$ are then given by (\ref{c17}).
Note that for the 3-bit model $G^*_{-+}=G_{ij}(2)$ and
$G^*_{0+}=G_{0+}(1)$, where $r=1$ and $r=2$ denote the nearest and next nearest
neighbor sites.
The relative importance of the pre- and postcollision pair correlations
$G_{ij}$ and $G^*_{ij}$ is best measured when they are normalized by the
single channel fluctuations,
$g_i=\langle (\delta n_i)^2 \rangle$, yielding the covariances\footnote{
In Ref.~\cite{Bussemaker92} the correlations $G_{ij}$ were normalized
by $f_i f_j$ rather than by $\sqrt{g_i g_j}$, as is done in this paper.
Only with the latter choice the self-dual symmetry is preserved.
}
\beq{e1}
	{\rm Cov}(i,j)   = \frac{G_{ij}}{\sqrt{g_i g_j}}, \qquad
	{\rm Cov}^*(i,j) = \frac{G^*_{ij}}{\sqrt{g_i g_j}}.
\eeq
Using the iterative scheme of section 4 we have calculated
these quantities for different choices of reduced density $f=\rho/3$,
lattice size $L$, and transition probabilities $\alpha, \beta, \gamma$
(for self-dual models only),
and compared the theoretical results with computer simulations.

\begin{figure}[h]
\centerline{
  \psfig{figure=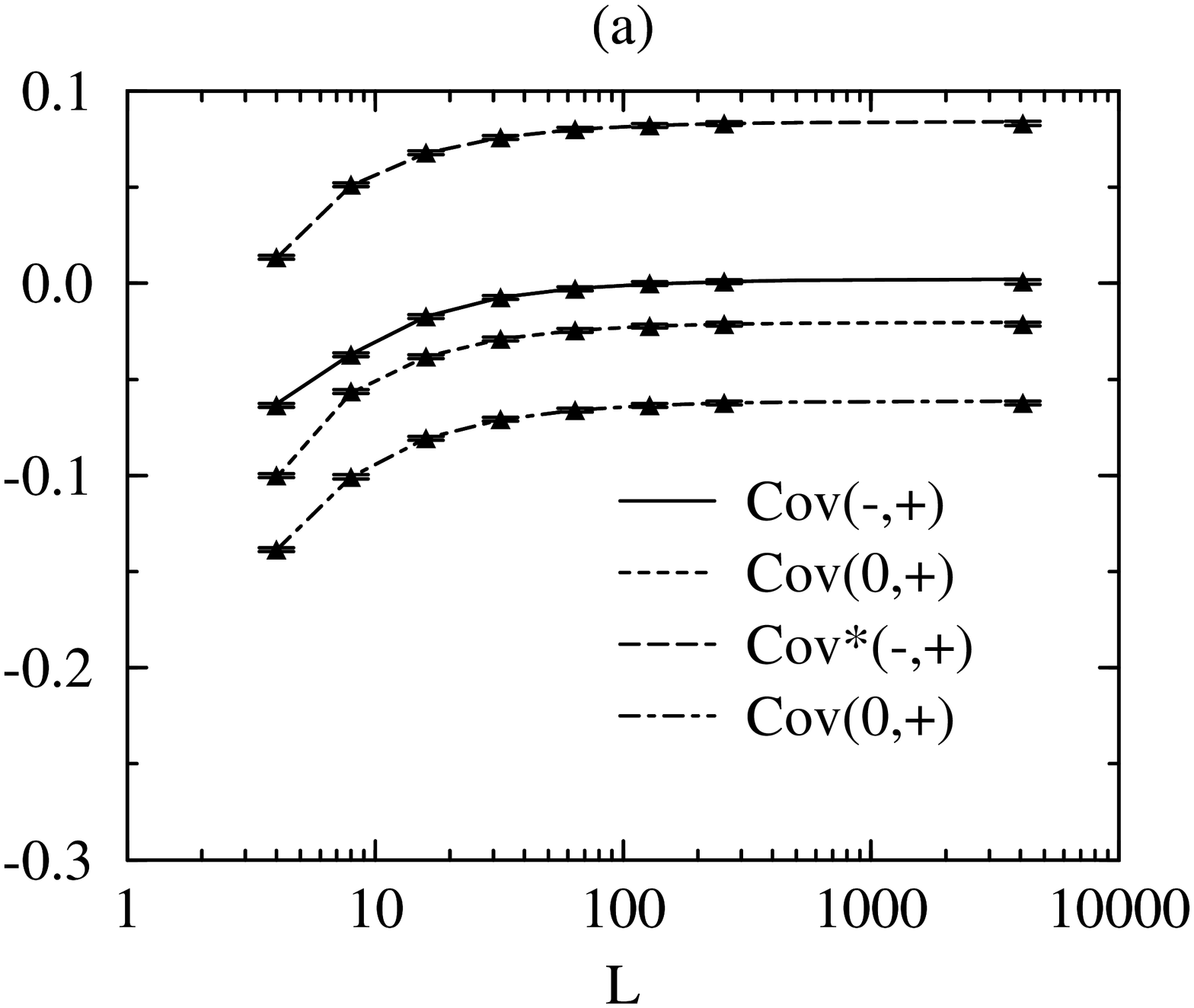,height=5cm,angle=270}
  \psfig{figure=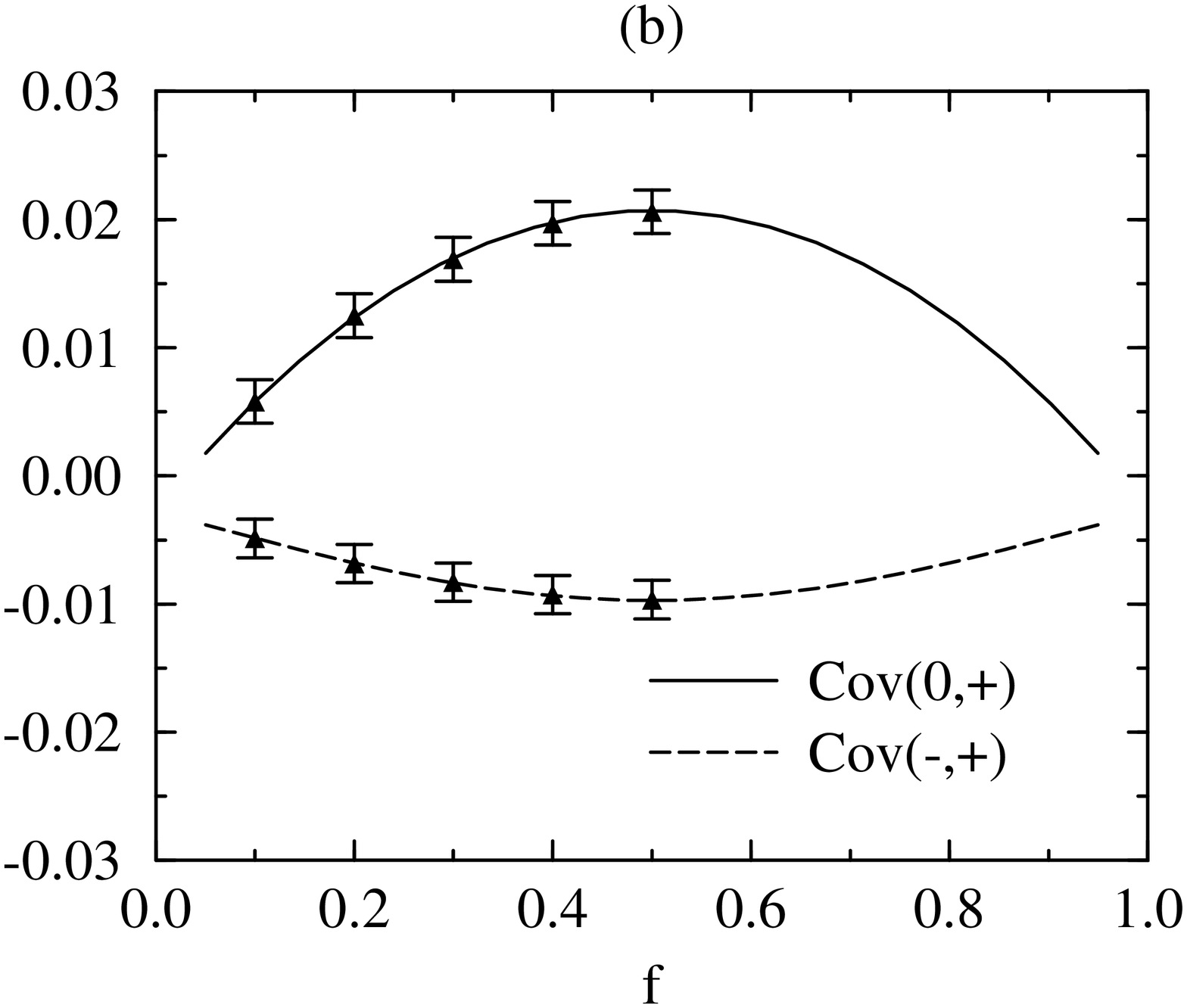,height=5cm,angle=270}
}
\centerline{
  \psfig{figure=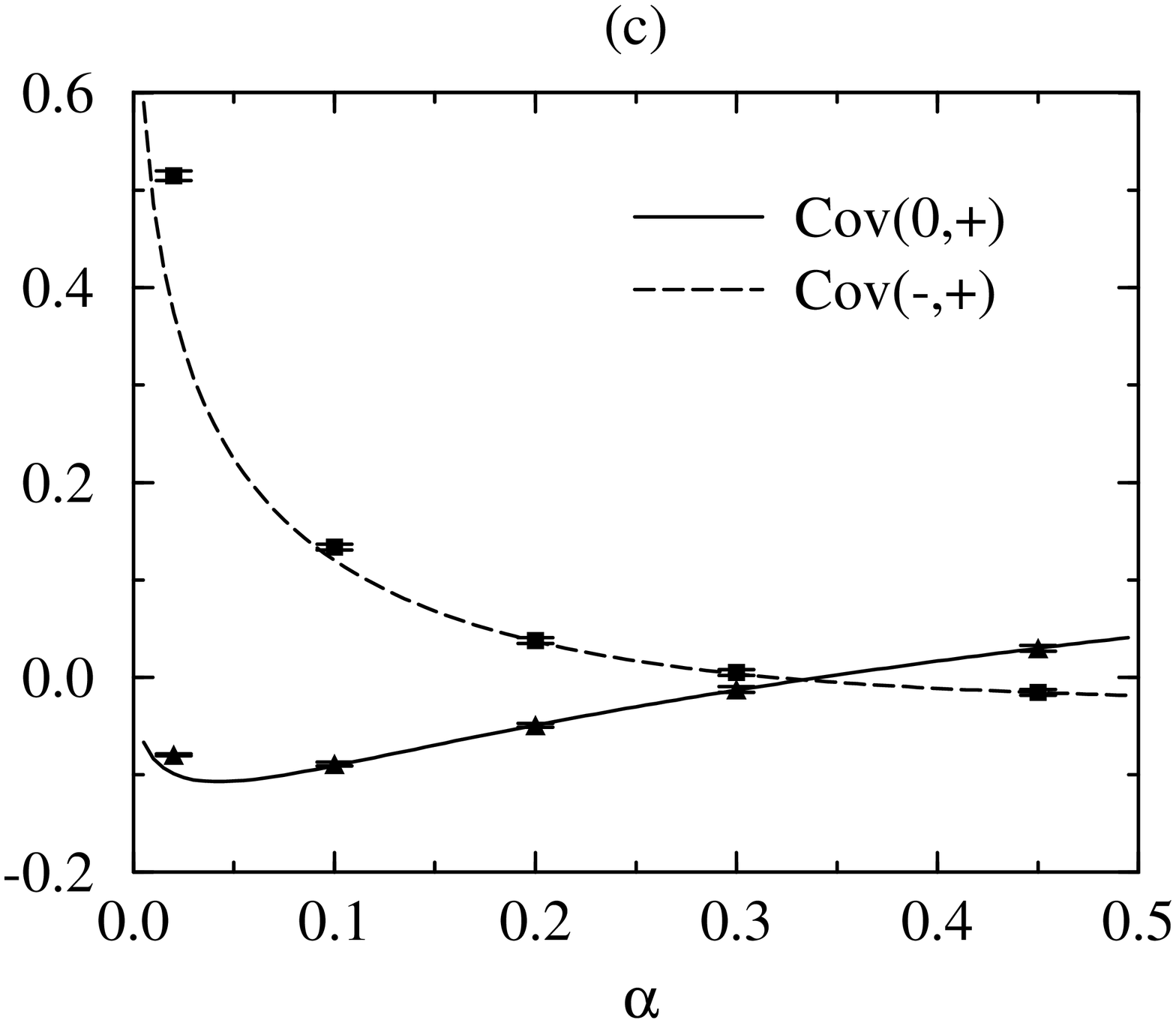,height=5cm,angle=270}
}
\caption{
Comparison of computer simulation results (symbols with error bars)
and self-consistent ring kinetic theory (lines) for the model of
Fig.~\protect\ref{fig1}.
On-node correlations, indicated as covariances, are plotted
(a) versus system size $L$, for transition probabilities $\alpha=0.4$,
    $\beta=0.5$, $\gamma=0$ and density $f=0.5$;
(b) versus density $f$, for $L=128$, $\alpha=0.5$, $\beta=0.4$, and $\gamma=0$;
(c) versus transition probability $\alpha$, for $f=0.5$, $L=128$, $\beta=0.33$,
    and $\gamma=0.5$.
Note the vanishing of correlations when $\alpha=\beta$ (detailed balance).
}
\label{fig2}
\end{figure}

In Fig.~\ref{fig2}a values of the pre- and postcollision on-node correlations,
obtained both from the numerical evaluation of the theory (lines) and from
computer simulations (symbols with error bars), are plotted as a function of
the system size $L$.
For small values of $L$ there are strong finite size effects,
which are quantitatively very well predicted by the theory.

It should be stressed that in a closed (microcanonical) system, with finite
$L$ and fixed $N$, correlations are always present between occupation numbers,
even in models {\em satisfying} semi-detailed balance.
This can be seen as follows: for a semi-detailed balance model with $V$ nodes
and $N=bfV$ particles we have if $(\bmr,\bmc_i) \neq (\bmr',\bmc_j)$ the
following relation,
$\langle n_i(\bmr) n_j(\bmr') \rangle = f [(N-1)/(bV-1)] < f^2$,
so that $G_{ij}(\bmr) = -{\cal O}(1/V) < 0$, independent of $\bmr$.

Fig.~\ref{fig2}b shows the typical dependence on the reduced density
$f=\rho/3$; the correlation functions are symmetric around $f=0.5$, as a
consequence of the imposed self-duality.
In Fig.~\ref{fig2}c the on-node correlations are plotted as a function of
$\alpha$, for fixed $\beta=0.33$ and $\gamma=0.5$ at
reduced density $f=0.5$ and $L=128$.
Note that when $\alpha=\beta$ all correlations vanish:
the model satisfies (semi)-detailed balance.

\begin{figure}[h]
\centerline{
  \psfig{figure=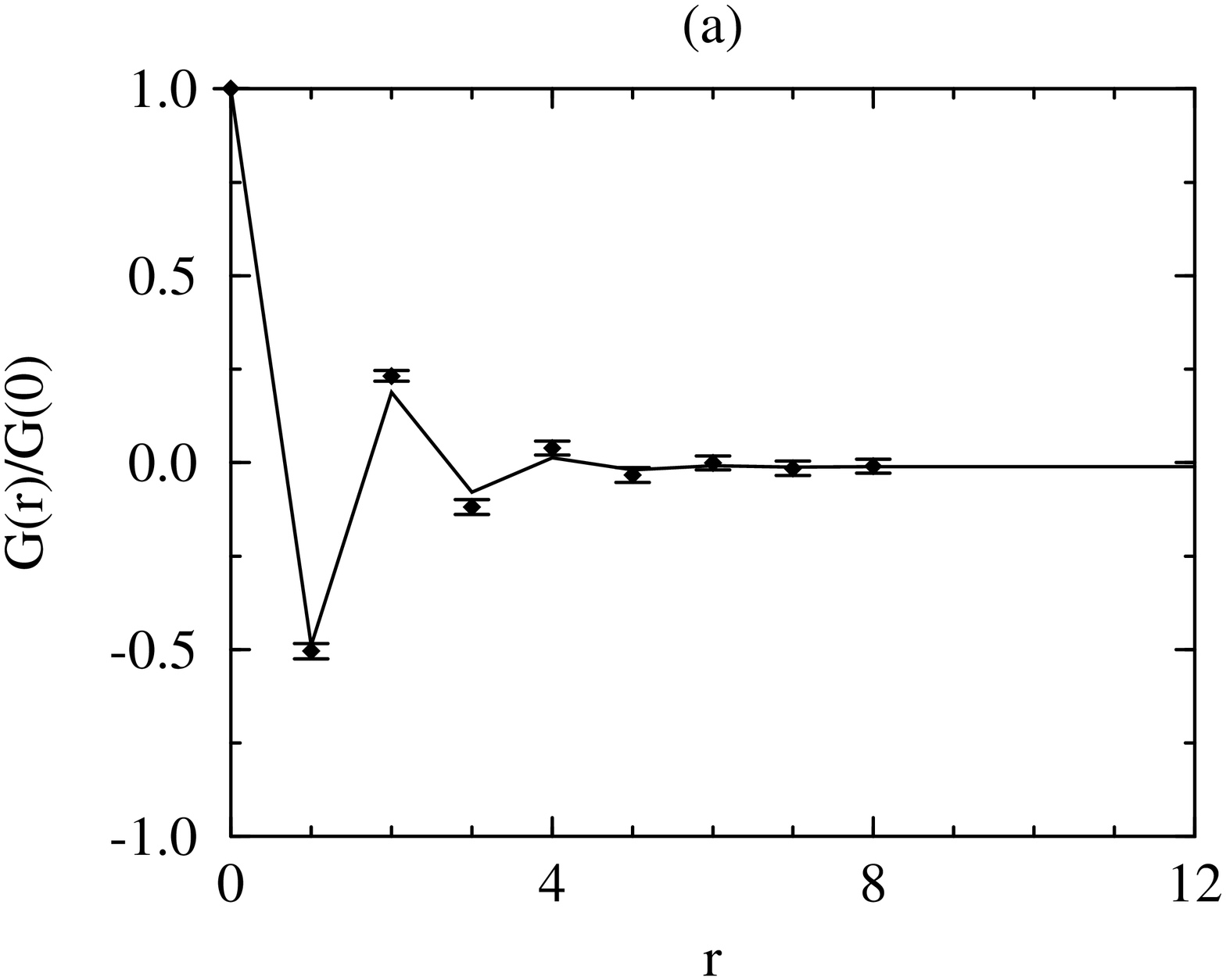,height=5cm,angle=270}
  \psfig{figure=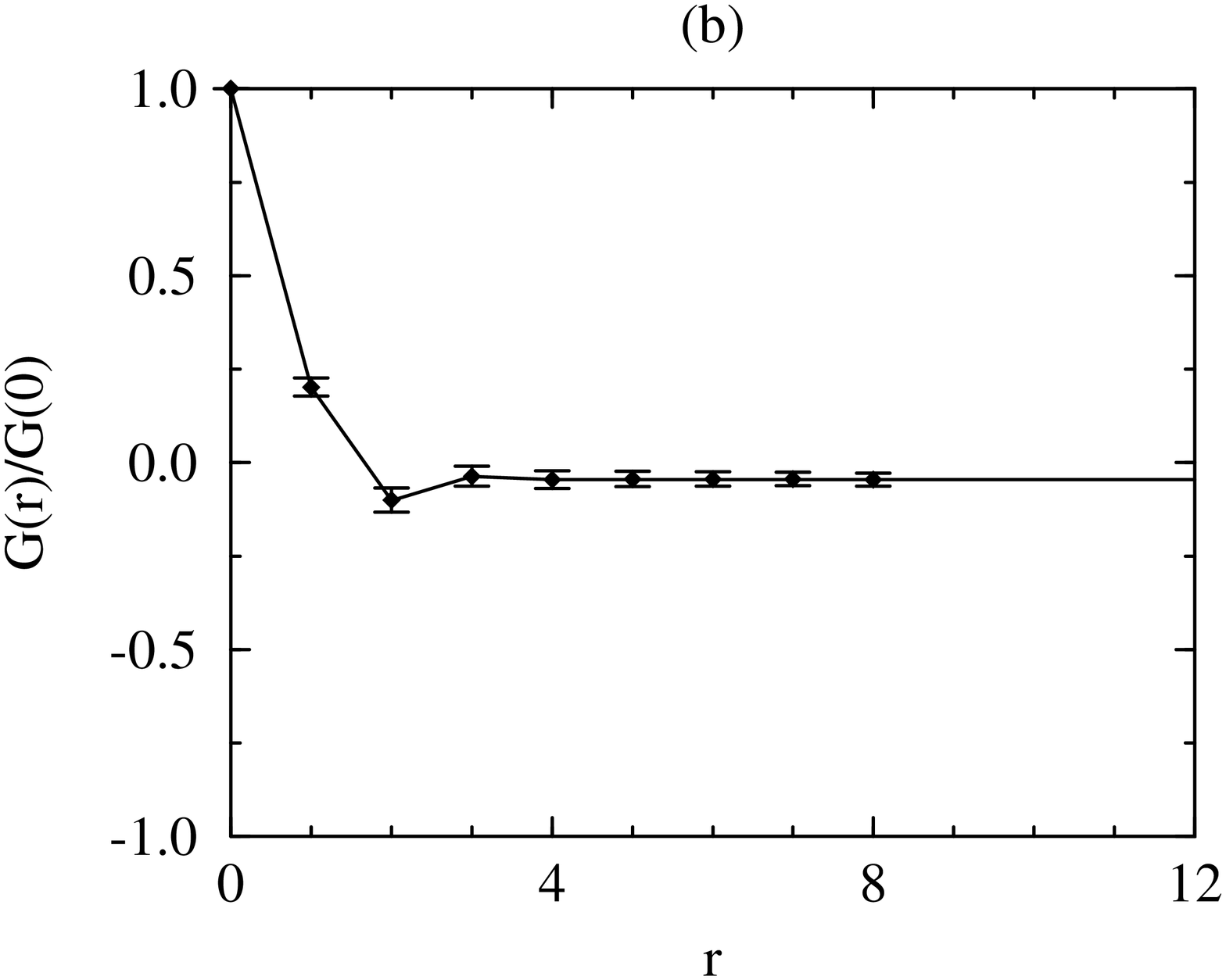,height=5cm,angle=270}
}
\caption{Off-node density-density correlation function
$G(r)=\sum_{ij}G_{ij}(r)$ plotted versus $r$,
for $f=0.5$, $L=256$, and two different
choices of transition probabilities:
(a) $\alpha=0.1$, $\beta=0.5$, $\gamma=0.5$, and
(b) $\alpha=0.5$, $\beta=0.1$, $\gamma=0.5$.
Simulations (symbols with error bars) compared with theoretical values
(solid line connecting points $r=0,1,2,..$).
}
\label{fig3}
\end{figure}

Since there is
no fundamental difference between on- and off-node correlations, and since in
addition the $L$-dependence of the correlations is well predicted by the
theory, we expect that our theory is capable of giving good predictions for
off-node correlations $G_{ij}(r)$ as well. This expectation is justified by
Fig.~\ref{fig3}a,b, where for fixed $f=0.5$, $L=128$ and two different sets of
transition probabilities, the density-density correlation function
$G(r)=\sum_{ij}G_{ij}(r)$ is plotted versus $r$.
Note that $G(r)$ is negative at large $r$, due to the finite size effects.

\begin{figure}[h]
\centerline{
  \psfig{figure=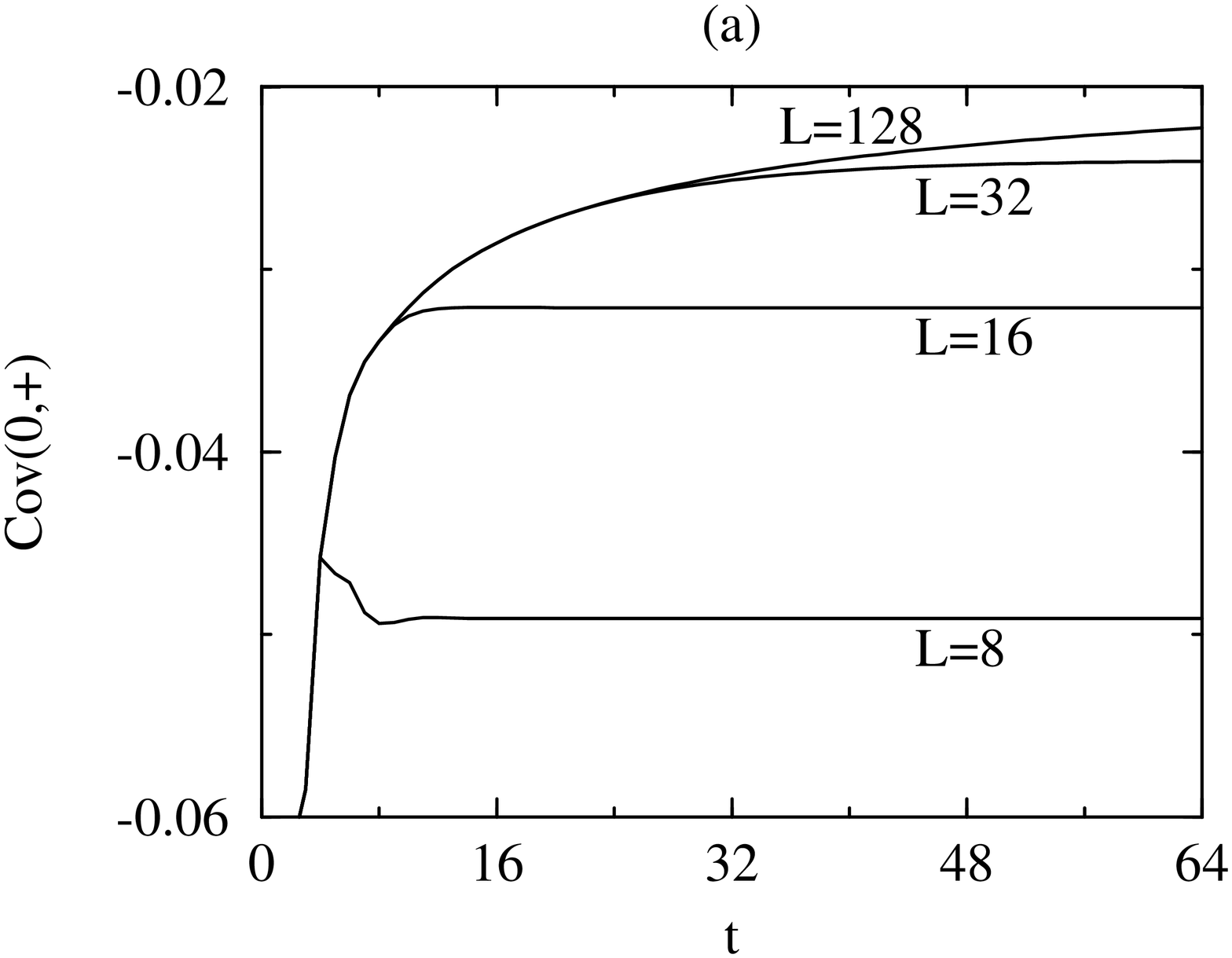,height=5cm,angle=270}
  \psfig{figure=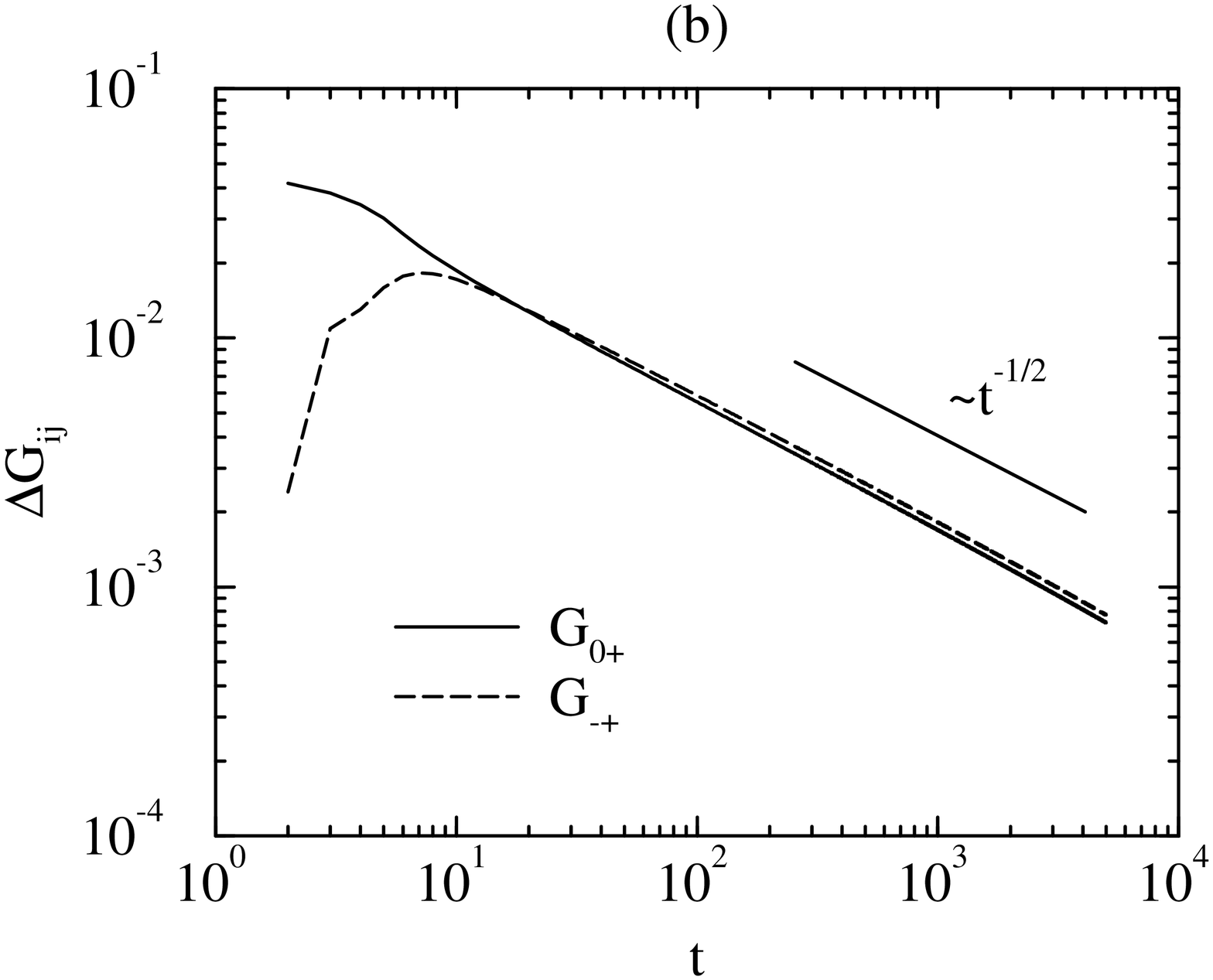,height=5cm,angle=270}
}
\caption{
Time-dependence of on-node correlations, calculated using ring kinetic
equation (no simulations), with $f=0.5$ and the transition probabilities of
Fig.~\protect\ref{fig2}a:
(a) short time behavior for various $L$;
(b) algebraic long time behavior, $\Delta G_{ij}\sim t^{-1/2}$, for $L=4096$.
}
\label{fig4}
\end{figure}

To study how the system approaches equilibrium in a spatially uniform
non-equilibrium state we have solved the time-dependent evolution equations
(\ref{b8}) and (\ref{b9}).  Figure~\ref{fig4}a shows how the typical
relaxation time increases with $L$.  Note that for $t<L/2$ the evolution is
independent of $L$, which can be explained by the observation that the shortest
time in which two opposite moving particles can meet through the periodic
boundary conditions is $t=L/2$; at earlier times there is no distinction
between a finite and an infinite system.  When $L=1024$ the system still has
not reached equilibrium after 1000 time steps.

For long times the approach to equilibrium is algebraic,
$G_{ij}(t)-G_{ij}(\infty) \equiv\Delta G_{ij}(t) \sim t^{-\alpha}$, with
an exponent $\alpha$ that depends on the dimensionality
and the type of collisional invariants of  the model.
Fig.~\ref{fig4}b shows that $\alpha = \frac{1}{2}$ in the case of the
interacting random walkers, where the number of particles is the only
collisional invariant.
In a separate publication we will analyze the exponents and amplitudes for
these algebraic tails in detail.

\subsection{Two-dimensional fluid-type model}
In a previous publication \cite{Bussemaker92} on-node correlations have been
studied in a 7-bit two-dimensional fluid-type LGA on a triangular
lattice, which violates (semi)-detailed balance.
Each node can contain a rest particle, and up to six moving
particles with velocities corresponding to nearest neighbor vectors.
Even after imposing that the collision rules conserve the lattice
symmetries, the model still has 20 independent transition probabilities.

Special choices of these probabilities, referred to as set \#$n$, have been
studied in Ref.~\cite{Bussemaker92}, where on-node correlations were measured
from computer simulations. For certain choices of transition probabilities
(e.g. set \#30) the correlations did not reach an equilibrium value, but
kept  growing in time.
This behavior could later be assigned to the existence of unstable sound modes,
driving a phase separation \cite{Bussemaker93}.  However, there
were other choices of transition probabilities (e.g. set \#13) that did
violate detailed balance while all modes remained stable.
In these cases a stationary state was
reached after a few hundred time steps.  In that paper only the on-node
postcollision correlations could be calculated in lowest approximation, i.e.
$C^*_{ij} = \Omega^{(2,0)}_{ij}(f^0)$
(see discussion below (\ref{c17})), but a
quantitative theory for the precollision correlations in equilibrium was
entirely lacking \cite{Bussemaker92}. The present paper provides the missing
theory.

\begin{figure}[h]
\centerline{
  \psfig{figure=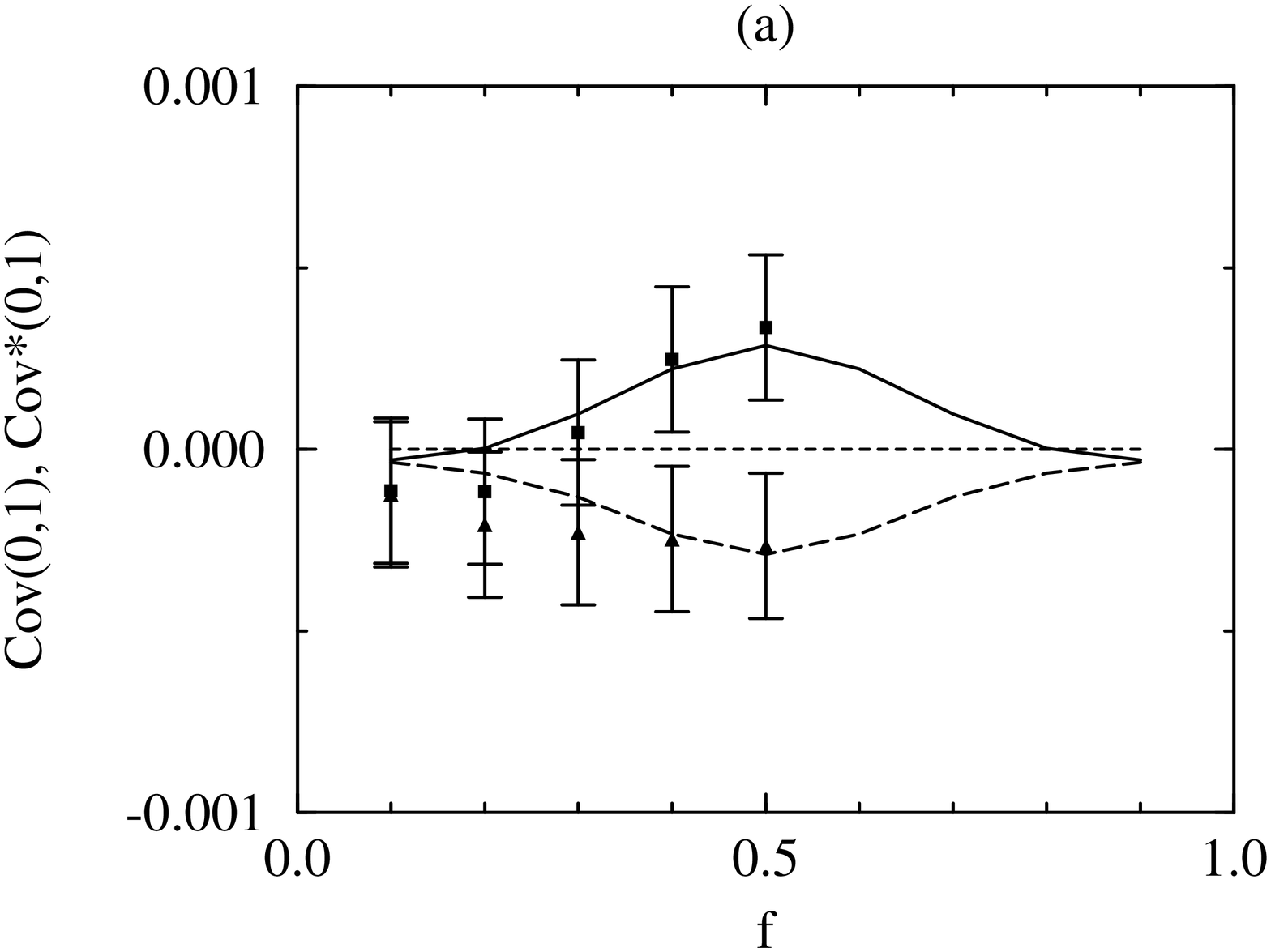,height=5cm,angle=270}
  \psfig{figure=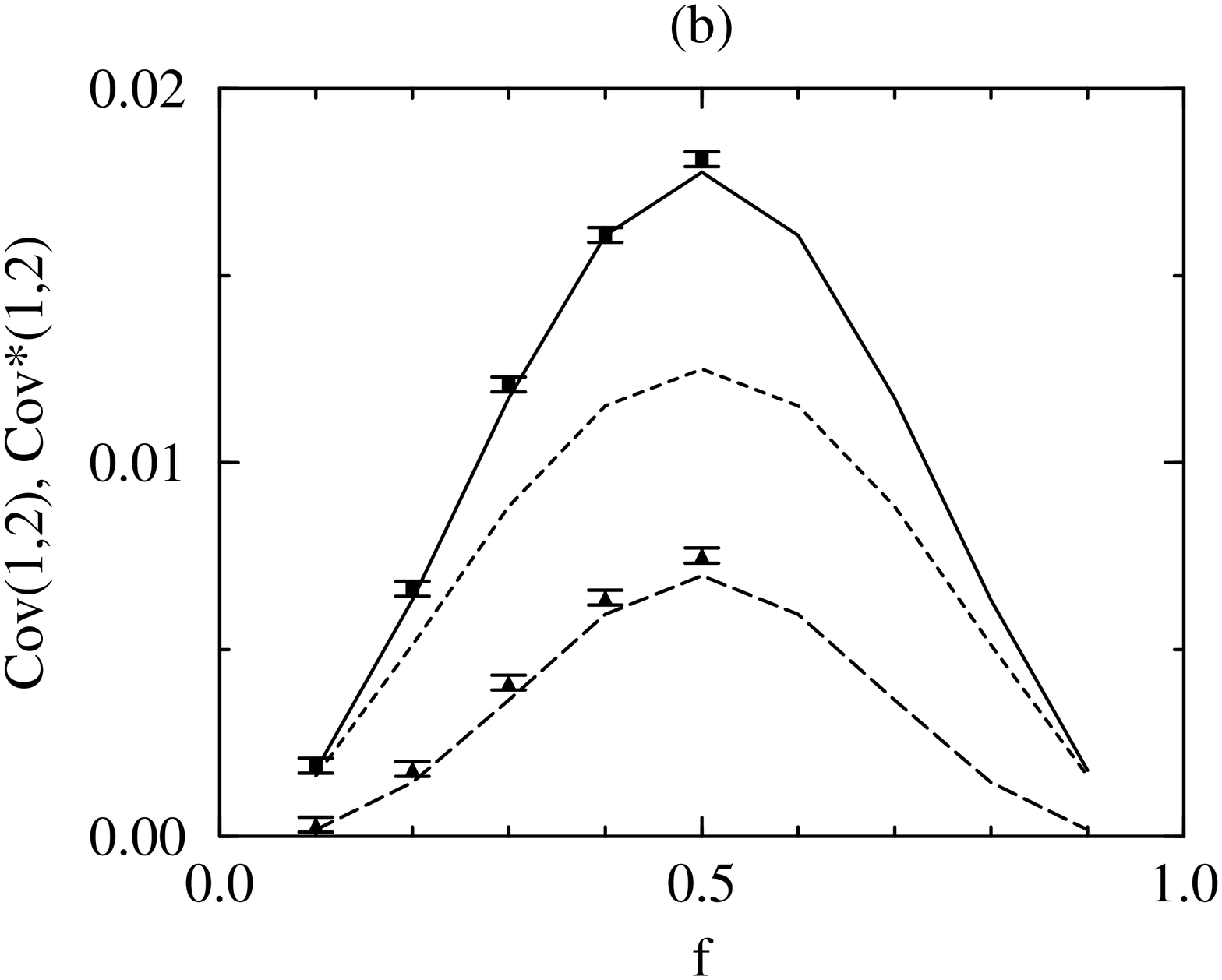,height=5cm,angle=270}
}
\centerline{
  \psfig{figure=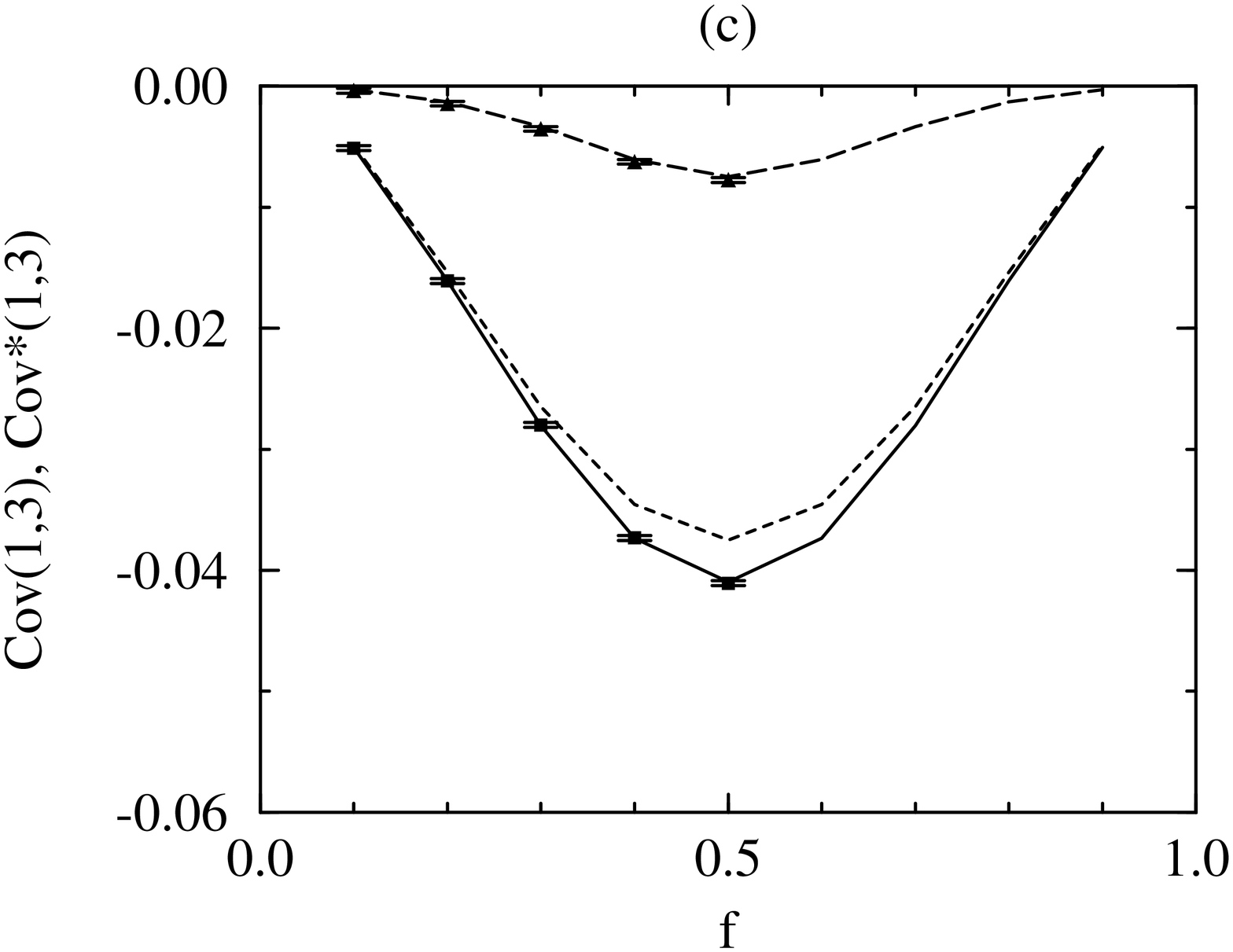,height=5cm,angle=270}
  \psfig{figure=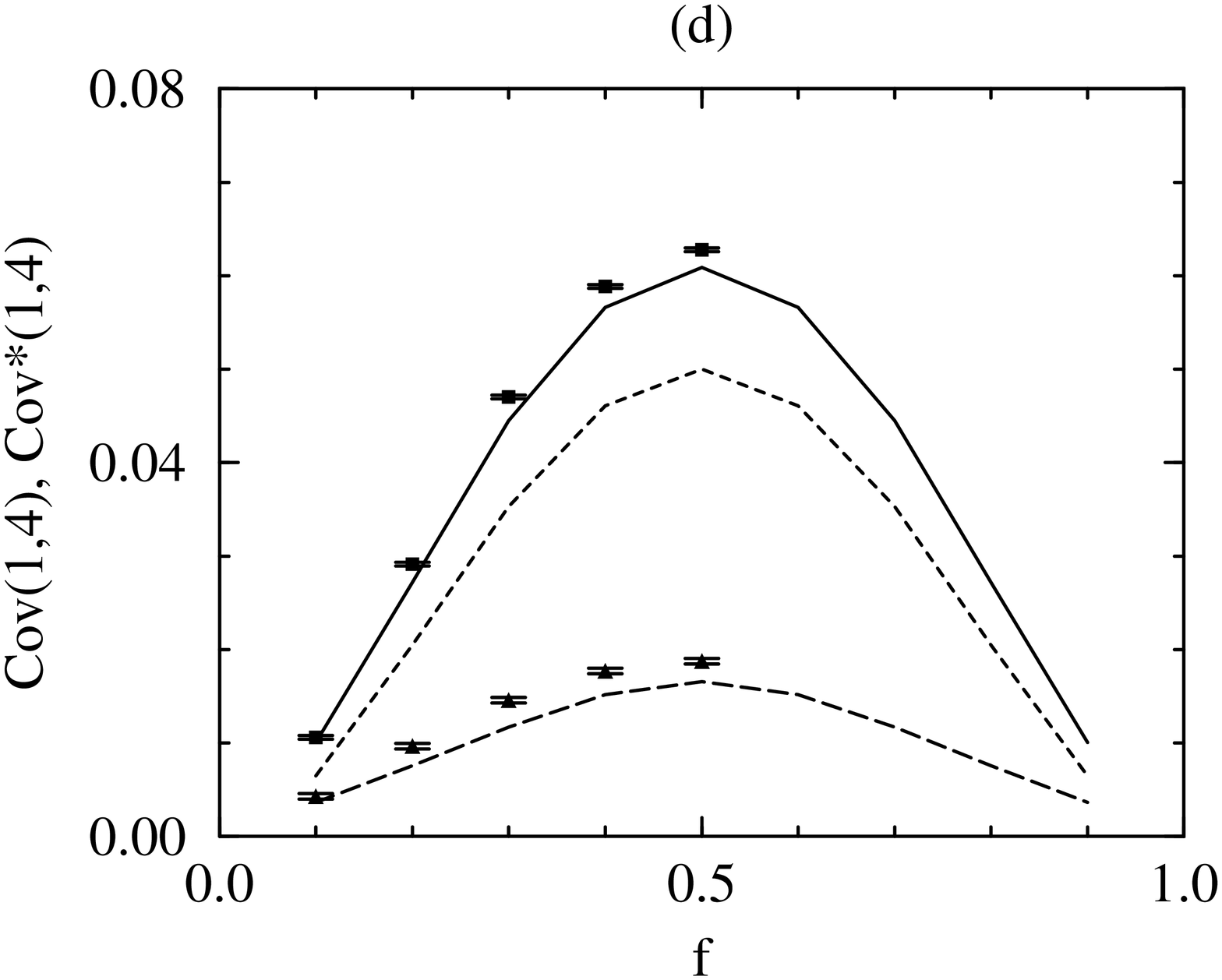,height=5cm,angle=270}
}
\caption{
Test of ring kinetic theory for the model of Ref.~\protect\cite{Bussemaker92},
using the transition probabilities of set \#13.
Simulation data for $L=64$ and $f=0.5$ (symbols with error bars),
compared with old theory for on-node postcollision correlations
of Ref.~\protect\cite{Bussemaker92} (dotted line), and with
ring kinetic theory for the on-node precollision (dashed line)
and postcollision (solid line) correlations.
}
\label{fig5}
\end{figure}

To test our theory for this non-detailed balance LGA we have chosen the set
\#13 of transition probabilities, and evaluated the ring kinetic theory. The
theoretical predictions for the correlation functions are compared with new
computer simulations of higher statistical accuracy, taking several hours of
CPU time on a RISC workstation for each data point.
In basic equilibrium, where the total momentum $\bmP$ vanishes, the
only independent elements of $C_{ij}$ are $(ij)=\{(01),(12),(13),(14)\}$;
all other elements are related to these four by lattice symmetries.
As set \#13 corresponds to a self-dual model, the correlations are again
symmetric around $f=\rho/7=0.5$.
Fig.~\ref{fig5}a-d shows the simulation results (symbols)
compared with the ring kinetic theory for precollision (solid line) and
postcollision (dashed line) correlations.
{}From this figure it is clear that the self-consistent ring kinetic theory
agrees very well with the simulations.
For comparison we have also plotted the functions $\Omega^{(2,0)}_{ij}(f^0)$
(dotted line) which represent the postcollision correlations created by
a single collision step acting on a factorized precollision state with
average occupations $f^0$ obtained from the nonlinear Boltzmann equation
(see discussion below eq.~\ref{b14}).
In an earlier paper \cite{Bussemaker92} we obtained
$C^*_{ij}=\Omega^{(2,0)}(f^0)$ as a simple estimate for the postcollision
correlations.

\setcounter{equation}{0}
\setcounter{section}{5}
\section{Conclusion}

We summarize the conclusions, consequences and possible applications as a set
of comments.

\noindent 1)
In the present paper, for the first time a theory is presented to calculate
distribution and correlation functions in lattice gas automata (LGA's) that
violate semi-detailed balance.
The theory of standard LGA's satisfying semi-detailed balance can be recovered
as a special case.

In section~2, starting from an open BBGKY hierarchy of coupled time evolution
equations for the $n$-particle distribution functions, we obtain closure by
neglecting correlations of third and higher order.
We are left with two coupled time evolution equations for the single
particle distribution function, $f_i(\bmr,t)$, and the pair correlation
function, $G_{ij}(\bmr,\bmr',t)$.
In section~3, to gain insight in the equilibrium state for LGA's violating
semi-detailed balance, we specialize our theory to a spatially uniform
stationary state, $f_i(\bmr,t)=f_i$ and
$G_{ij}(\bmr,\bmr',t)=G_{ij}(\bmr-\bmr')$.
For given $f_i$, the on-node correlations $G_{ij}$ are uniquely determined by a
linear equation: the so-called ring equation (\ref{c9c}).

\noindent 2)
The ring operator $R$ defined in (\ref{c16c}) is identical to the
one introduced for calculating long time tails of time correlation functions
\cite{Kirkpatrick91,Brito92} and corrections to Boltzmann transport
coefficients \cite{Velzen93} in LGA's satisfying semi-detailed balance.
One can show that linearization of the theory in this paper around the
Gibbsian equilibrium state for standard LGA's reproduces the results of
Refs.~\cite{Kirkpatrick91,Brito92,Velzen93}.
Using the spectral decomposition method discussed in appendix~B -- rather
than the method discussed in Ref.~\cite{Velzen93} -- in combination with
a Gaussian integration scheme for performing the $\bmq$-integral in
(\ref{c16c}), the ring operator can efficiently be evaluated numerically
for large system sizes.

\noindent 3)
In section~4 we provide a numerical scheme for finding, by self-consistent
iteration, a stationary solution $\{f_i,G_{ij}\}$ that is `close' to
the solution of the nonlinear Boltzmann equation (where $G_{ij} \equiv 0$).
Is the equilibrium solution found by this scheme unique?
We note that even though the ring equation uniquely determines $G_{ij}$
for given $f_i$, it is highly nonlinear in $f_i$.
Therefore it can in principle not be excluded that more than one spatially
uniform stationary solution, or equilibrium state, exists.

\noindent 4)
It is well known that the violation of semi-detailed balance in LGA's may
lead to spatial instability of the uniform state and subsequent pattern
formation \cite{LGA-patt}.
Computer simulation studies of a two-dimensional LGA violating semi-detailed
balance, where the presence of unstable sound modes leads to phase separation,
have indicated that a finite system with periodic boundary conditions randomly
`chooses' between different spatially inhomogeneous stationary states,
corresponding to limit cycles with a `continuously' broken symmetry
\cite{Bussemaker93}.

\noindent 5)
To test our theory we have applied it to two different LGA's, both violating
(semi)-detailed balance: a one-dimensional model of interacting random walkers,
and a two-dimensional fluid-type LGA defined on a triangular lattice.
For the first model we have compared theory with simulations for various
densities, transition probabilities and system sizes. In all cases there is
remarkable agreement. Even the finite size effects occurring for small system
sizes are well predicted by the theory.
The small deviations that occur in a few cases are probably due to the fact
that
third and higher order correlations are systematically neglected.

\noindent 6)
The second model was already considered in an earlier publication
\cite{Bussemaker92}.  In that paper a simple theoretical estimate
was given for the postcollision on-node correlations,
$C^*_{ij} = \Omega^{(2,0)}_{ij}$, in reasonable agreement with computer
simulations. However, theoretical predictions for the precollision on-node
correlations, and for off-node correlations were totally lacking.
The present paper provides the missing theory, and also shows the excellent
agreement between theory and simulations for this two-dimensional fluid model.

\noindent 7)
As $t\rightarrow\infty$ a stationary state is approached. Due to the existence
of local conservation laws the approach to this stationary state is algebraic,
$\sim t^{-\alpha}$, with an exponent $\alpha$ that depends on the
dimensionality and on the type of conservation laws.
There is an intimate connection between the algebraic tails in the approach
towards the correlated equilibrium state mentioned here, and the well-known
long time tails in the decay of the velocity autocorrelation function
\cite{Frenkel89,Ernst90}. More generally, several ideas from the field of
classical kinetic theory can be applied to the equations derived in this
paper, which will be discussed in a separate publication.

\noindent 8)
In Ref.~\cite{Dufty94} the Boltzmann-Langevin equation for LGA's was discussed.
This phenomenological equation requires specification of the correlation
function of the random noise sources. To what extent does the
fluctuation-dissipation theorem enable us to relate the noise strength to
the phenomenological damping or transport coefficients?
The fluctuation-dissipation theorem requires the stationary distribution
and correlation functions as input.
If one does not know this distribution function a priori the
fluctuation-dissipation theorem is an empty statement.
For LGA's that {\em violate} (semi)-detailed balance, one may therefore also
summarize the implications of the previous discussion by stating that the
standard form of the {\em fluctuation-dissipation} theorem for Langevin's
equation, relating the noise strengths to the dissipation coefficients,
is {\em no longer valid}, because the stationary distribution differs from
the Gibbs distribution.

\noindent 9)
It is appropriate to give some comments on the validity of the generalized
Boltzmann equation and ring equation derived in this paper.
In real fluids the validity of the cluster expansion is restricted to dilute
systems, where the density acts as a small expansion parameter.
Here there is no such small parameter and the approximations are less
controlled.
To understand the great success of the present mean-field theory, it is
useful to compare the lattice Boltzmann equation with the revised Enskog
theory for hard sphere fluids \cite{deSchepper}.
Both theories describe non-equilibrium phenomena reasonably well over the
whole density range. The reason for this seems to be that the collision term
for the Revised Enskog Theory (resp.\ Lattice Boltzmann equation) contains the
{\em static correlations} imposed by the hard core exclusion (resp.\ Fermi
exclusion), which are the origin for the strong density dependence of the
theory.
The present ring kinetic theory equation also contains -- through the
expansion coefficients $\Omega$ -- all static correlations between colliding
particles imposed by the Fermi exclusion. This might explain the success
in predicting the density dependence of the correlation functions.


\appendix
\renewcommand{\theequation}{A.\arabic{equation}}
\setcounter{equation}{0}
\section*{Appendix A}
In this appendix the cluster expansion of the hierarchy equations is
derived.
We start by substituting $n_i=f_i+\delta n_i$ into identity (\ref{a6})
and obtain,
\beq{b1}
  (f_i+\delta n_j)^{s_j} (1-f_j - \delta n_j)^{1-s_j}
	= f_j^{s_j}(1-f_j)^{1-s_j}
	  \left[ 1 + \frac{\delta s_j \delta n_j}{g_j} \right],
\eeq
with $g_j = f_j(1-f_j)$.
With the help of this identity we can expand the single node distribution
$p(s,\bmr,t)$, defined in (\ref{a10}),
\beq{b2b}
  p(s,\bmr,t)	= F(s,\bmr,t) \left\{1+\sum_{k<\ell}
	\frac{\delta s_k(\bmr,t) \delta s_l(\bmr,t)}{g_k(\bmr,t) g_l(\bmr,t)}
	    G_{k\ell}(\bmr,\bmr,t) + \ldots \right\},
\eeq
where $F(s,\bmr,t)$ is the completely factorized single node
distribution function defined in (\ref{b3}).
Similarly we can expand the two-node distribution function in (\ref{a10a}) as
\beqa{b4}
  \lefteqn{ p^{(2)}(s,\bmr,s',\bmr',t) - p(s,\bmr,t) p(s',\bmr',t) = }
  \nonumber \\
  & & F(s,\bmr,t) F(s',\bmr',t) \sum_{k<\ell} \frac{\delta
s_k(\bmr)}{g_k(\bmr,t)}
  \frac{\delta s_l(\bmr')}{g_l(\bmr',t)} G_{k\ell}(\bmr,\bmr',t) + \ldots .
\eeqa
Using the normalization (\ref{a1}) of $A_{s \sigma}$
we rewrite the time evolution equation (\ref{a6a}) for the single particle
distribution function $f_i(\bmr,t)$ as,
\beq{m1}
  f_i(\bmr+\bmc_i,t+1) - f_i(\bmr,t)  =
 	\sum_{s \sigma} (\sigma_i - s_i) A_{s \sigma} p(s,\bmr,t).
\eeq
Insertion of (\ref{b2b}) yields
\beq{m2}
  f_i(\bmr+\bmc_i,t+1) - f_i(\bmr,t)
	\simeq \Omega^{(1,0)}_i(f) + \sum_{k<\ell} \Omega^{(1,2)}_{i,k\ell}(f)
	    G_{k\ell}(\bmr,\bmr,t).
\eeq
The expansion coefficients $\Omega$ are given by
\beqa{m3}
 \Omega^{(1,0)}_i(f) &=&  \sum_{s \sigma} (\delta \sigma_i -
 	\delta s_i) A_{s \sigma} F(s)  \nonumber \\
 \Omega^{(1,1)}_{ij}(f) &=&
	\sum_{s \sigma} (\delta\sigma_i - \delta s_i)
	A_{s \sigma} F(s) \frac{\delta s_j}{g_j} \nonumber \\
 \Omega^{(1,2)}_{i,k\ell}(f) &=&
	\sum_{s \sigma} (\delta\sigma_i - \delta s_i)
	A_{s \sigma} F(s) \frac{\delta s_k \delta s_\ell}{g_k g_\ell},
\eeqa
where $k \neq \ell$ and $F(s) = F(s,\bmr,t)$ is the {\em completely factorized}
single node distribution function defined in (\ref{b3}).
Note that the $\Omega$'s depend on $\bmr$ and $t$ through $f(\bmr,t)$.
They are generated by the Taylor expansion of the nonlinear Boltzmann collision
operator, $\Omega^{(1,0)}_i(f + \delta f)$, in powers of $\delta f$.

To derive an equation of motion for $G_{ij}(\bmr,\bmr',t)$, defined in
(\ref{a11}), we start from
\beqa{m4m}
  G_{ij}(\bmr +\bmc_i,\bmr'+ \bmc_j,t+1) &=&
  	f^{(2)}_{ij}(\bmr +\bmc_i,\bmr'+ \bmc_j,t+1) - \\
	&& \hspace{-40mm} f_i(\bmr+\bmc_i,t+1) f_j(\bmr'+\bmc_j,t+1)
	\nonumber
\eeqa
and use (\ref{a6a}) together with (\ref{a8a}) to obtain
\beq{m4}
 G_{ij}(\bmr +\bmc_i,\bmr'+ \bmc_j,t+1) =
  K_{ij}(\bmr,\bmr',t) + \delta (\bmr, \bmr') B_{ij} (\bmr,t) ,
\eeq
with
\beq{m5}
 K_{ij}(\bmr,\bmr',t) =
 \sum_{s \sigma} \sum_{s' \sigma'}
 \delta\sigma_i \delta\sigma'_j A_{s \sigma} A_{s' \sigma'}
 \left\{ p^{(2)}(s,\bmr,s',\bmr',t) - p(s,\bmr,t) p(s',\bmr',t) \right\}.
\eeq
and the on-node source term,
\beq{m6}
  B_{ij}(\bmr,t) = B^{(1)}_{ij}(\bmr,t) + B^{(2)}_{ij}(\bmr,t)
			+ B^{(3)}_{ij}(\bmr,t),
\eeq
consisting of three parts,
\beqa{}
  B^{(1)}_{ij}(\bmr,t) &=&
	\sum_{s \sigma} \delta\sigma_i \delta\sigma_j
	A_{s \sigma} p(s,\bmr,t) \nonumber\\
	&=& G_{ij} (\bmr, \bmr, t) + \sum_{s \sigma}
	(\delta\sigma_i \delta\sigma_j - \delta s_i \delta s_j)
	A_{s \sigma} p(s) \nonumber\\
	\nonumber\\
  B^{(2)}_{ij}(\bmr,t) &=&
	- \sum_{s \sigma} \sum_{s' \sigma'} \delta\sigma_i \delta\sigma'_j
	A_{s \sigma} A_{s' \sigma'}
	[p^{(2)}(s,\bmr,s',\bmr,t) - p(s,\bmr,t)p(s',\bmr,t)] \nonumber\\
  B^{(3)}_{ij}(\bmr,t) &=&
	- \sum_{s \sigma} \sum_{s' \sigma'} \delta\sigma_i \delta\sigma'_j
	A_{s \sigma} A_{s' \sigma'} p(s,\bmr,t) p(s',\bmr,t).
\eeqa
The cluster expansion of $K_{ij}$ follows from (\ref{b4}) with the result
\beq{m7}
  K_{ij}(\bmr,\bmr',t) \simeq \omega_{ij,k\ell}(f,f') G_{k\ell}(\bmr,\bmr',t),
\eeq
where the pair collision operator is defined as
\beqa{m8}
  \omega_{ij,k\ell}(f,f') &=&
	\sum_{s \sigma s' {\sigma}^\prime}
	\delta\sigma_i \delta\sigma^\prime_j
	A_{s \sigma} A_{s' \sigma'}
	F(s) F(s') \frac{\delta s_k \delta s'_\ell}{g_k g_\ell}
	\nonumber \\
	&=& 	\{1 + \Omega^{(1,1)}(f(\bmr,t))\}_{ik}
		\{1 + \Omega^{(1,1)}(f(\bmr',t))\}_{j\ell}.
\eeqa
To identify $\Omega^{(1,1)}(f)$ with the symbol introduced in (\ref{m3}) one
needs to verify that the term containing $\delta s_i \delta s_j$ equals
$\delta_{ij}$.
For the three parts of $B_{ij}(\bmr,t)$ we obtain
\beqa{m10}
  B_{ij}^{(1)}(\bmr,t) &\simeq&
	G_{ij} (\bmr, \bmr, t) + \Omega^{(2,0)}_{ij}(f)
	+ \sum_{k<\ell} \Omega^{(2,2)}_{ij,k\ell}(f) G_{k\ell}(\bmr,\bmr,t)
	\nonumber\\
  B^{(2)}_{ij} (\bmr,t) &\simeq&
	-\omega_{ij,k\ell}(f,f) G_{k\ell}(\bmr,\bmr,t) \\
  B^{(3)}_{ij} (\bmr,t) &\simeq&
	-\Omega^{(1,0)}_i(f) \Omega^{(1,0)}_j(f)  \nonumber\\
	&& \mbox{} - \sum_{k<\ell} \left\{
	\Omega^{(1,0)}_i(f) \Omega^{(1,2)}_{j,k\ell}(f)
	+ \Omega^{(1,0)}_j(f) \Omega^{(1,2)}_{i,k\ell}(f)
	\right \} G_{k\ell}(\bmr,\bmr,t). \nonumber
\eeqa
We have introduced two more expansion coefficients, where $k \neq \ell$,
\beqa{m13}
  \Omega^{(2,0)}_{ij}(f) &=&
	\sum_{s \sigma} (\delta\sigma_i \delta\sigma_j - \delta s_i \delta s_j)
	A_{s \sigma} F(s) \nonumber \\
  \Omega^{(2,2)}_{ij,k\ell}(f) &=&
	\sum_{s \sigma} (\delta\sigma_i \delta\sigma_j - \delta s_i \delta s_j)
	A_{s \sigma} F(s) \frac{\delta s_k \delta s_\ell}{g_k g_\ell}.
\eeqa

\appendix
\renewcommand{\theequation}{B.\arabic{equation}}
\setcounter{equation}{0}
\section*{Appendix B}

In this appendix we discuss the formal solution of the ring
equation (\ref{c4}),
\beq{c6a}
  \hatG(\bmq) = \frac{1}{1-s(\bmq)\omega} s(\bmq) B.
\eeq
Using a bi-orthogonal set of right and left eigenvectors,
$\tilde\chi_\alpha(\bmq)$ and
$\chi_\alpha(\bmq)$, and eigenvalues
$\lambda_\alpha(\bmq)$ of the two-particle
propagator $\gamma(\bmq)=s(\bmq)\omega$,
we can write $(1-s(\bmq)\omega)^{-1}$ as a spectral decomposition,
\beq{c6}
  \frac{1}{1-s(\bmq)\omega} = \sum_\alpha | \tilde\chi_\alpha (\bmq)
   \rangle \; \frac{1}{1-\lambda_\alpha(\bmq)} \; \langle \chi_\alpha(\bmq) |,
\eeq
provided that $\lambda_\alpha(\bmq) \neq 1$.
There is however a complication if, for one or more reciprocal lattice vectors
$\bmq_n$ belonging to a set $\{\bmq_n\}$, there are eigenmodes $\alpha$
for which $\lambda_\alpha(\bmq_n)=1$.  The operator $(1-s(\bmq_n\omega))$
then has a null space, spanned by the eigenvectors for which
$\lambda_\alpha(\bmq_n)=1$. The set of points $\{\bmq_n\}$ contains the origin,
corresponding to the standard conservation laws, and the centers of the facets
of the Wigner-Seitz cell, corresponding to the staggered invariants
\cite{Das93}.

We can now formulate a necessary {\em solubility condition}, which requires
that the inhomogeneous term in (\ref{c4}) be orthogonal to the null
space, i.e.
\beq{c7}
  \langle \chi_\alpha(\bmq_n) \: | s(\bmq_n) B \rangle = 0
\eeq
for the modes $(\alpha,n)$ that have $\lambda_\alpha(\bmq_n)=1$.
It can be verified that this condition is indeed satisfied
as a consequence of (\ref{b13}), i.e.\ the local conservation laws,
since both for the standard conserved quantities ${\cal H}$ and for
the staggered invariants we have $s(\bmq_n)=1$.

After establishing that (\ref{c4}) is soluble, we observe that its
solution is {\em not unique} in the set of points $\{\bmq_n\}$,  where
it is given by
\beq{c8}
  \hatG(\bmq_n) = Q_n \frac{1}{1-\omega} B
    + {\sum_\alpha}^\prime {\cal E}_{n\alpha} \tilde\chi_\alpha(\bmq_n),
\eeq
where the relation $s(\bmq_n)=1$ has been used. The coefficients
${\cal E}_{n\alpha}$ are undetermined, and the prime on the sum
indicates that the summation is restricted to eigenmodes $\alpha$ for
which $\lambda_\alpha=1$.
The projector $Q_n$, defined as
\beq{c9}
  Q_n = 1 - P_n = 1 - {\sum_\alpha}^\prime |
  \tilde\chi_\alpha(\bmq_n) \rangle  \langle \chi_\alpha(\bmq_n)|,
\eeq
is introduced to make explicit that the inverse of $(1 - s(\bmq_n) \omega)$
is only defined in the
orthogonal complement of the null space of $(1-s(\bmq_n)\omega)$, where
$\lambda_\alpha \neq 1$. The second term lies in the null space, i.e.
\beq{c10}
P_n \hatG(\bmq_n) = {\sum_\alpha}^\prime {\cal E}_{n \alpha}
\tilde{\chi}_\alpha (\bmq_n).
\eeq
The coefficients ${\cal E}_{n \alpha}$ are in fact equal to the covariance of
the fluctuations of the (standard and spurious) global invariants.  For
instance, for the standard invariants ${\cal H}=\{N,\bmP,...\}$, associated
with
$\bmq=\bmzero$, the eigenmode  $\chi_\alpha(\bmzero)=a_\lambda a_\mu$ is a
product of collisional invariants. Then
\beq{c11}
  {\cal E}_{n \alpha} \equiv {\cal E}_{\lambda\mu}
    = \langle a_\lambda a_\mu | \hatG(\bmzero) \rangle
    = \sum_\bmr \sum_{ij} a_{\lambda i} a_{\mu j} G_{ij}(\bmr)
    = \frac{1}{V} \langle \delta {\cal H}_\lambda \delta{\cal H}_\mu \rangle.
\eeq
Therefore the coefficients ${\cal E}_{n \alpha}$ are constant in time,
and can be calculated from the initial ensemble at $t=0$.
In the calculations of this paper we assume that the initial ensemble is
prepared {\em microcanonically}, i.e. with all global (standard and spurious)
invariants fixed, from which it follows that ${\cal E}_{n \alpha}=0$.

In the previous discussion we have explained the structure of the formal
solution $\hatG (\bmq)$ of (\ref{c4}). However,
this is not yet a solution of
the ring equation, or even a closed equation for  $\hatG (\bmq)$, because the
inhomogeneous term $B_{ij}$ is given in terms of the on-node distribution
function $G_{ij}$. It is related to  $\hatG_{ij}(\bmq)$ by
\beq{c13}
  G_{ij} = \frac{1}{V} \sum_\bmq \hatG_{ij}(\bmq),
\eeq
where the $\bmq$--summation runs over the first Brioullin zone
in the reciprocal lattice $\cal L^*$.

Combination of (\ref{c13}), (\ref{c8}), (\ref{c10}) and (\ref{c6a})
gives the following closed equation for the on-node correlation matrix $G$,
\beq{c14}
  G = R B + \frac{1}{V} \sum_n P_n \hatG(\bmq_n)
\eeq
with $B$ given in (\ref{c6c}) and the ring operator $R$ is
according to (\ref{c6}) and (\ref{c8}) defined as
\beqa{c15}
  R &=& \frac{1}{V} \sum_{\bmq \not\in \{\bmq_n\}} \frac{1}{1-
  s(\bmq)\omega} s(\bmq) +
  \frac{1}{V} \sum_n Q_n \frac{1}{1-\omega} \nonumber \\
  &\equiv& \frac{1}{V} {\sum_{\bmq}}^* \frac{1}{1-
  s(\bmq)\omega} s(\bmq),
\eeqa
where $s(\bmq_n) =1 $ has been used. The asterisk on the summation sign is
defined through the second equality.

As the diagonal part $G_d$ of the on-node correlation function is known for
given $f^0_i$, it is convenient to introduce the {\em excess correlation}
function $C= G -G_d$, and to absorb the contribution from $G_d$ into a known
inhomogeneous term. To do so,
we observe that $\Omega^{(2,2)} G_d=0$ on account of (\ref{m13}), and
we calculate
\beq{c18}
  R (1-\omega) G_d = \frac{1}{V} {\sum_{\bmq}}^*
	\frac{1}{1-s(\bmq)\omega} s(\bmq) (1-\omega) G_d
= \frac{1}{V} {\sum_{\bmq}}^* G_d .
\eeq
In the second equality we have used  the relation
$s(\bmq)G_d=G_d$ because $s_{ii}(\bmq)=1$.
Equation (\ref{c14}) yields then  for the excess
correlation function,
\beq{c19}
  C = R \{ \Omega^{(2,2)} + (1-\omega) \} C + R \Omega^{(2,0)} + J,
\eeq
with the finite size correction term $J$ given by,
\beqa{c20}
J &=& -G_d +  \frac{1}{V} {\sum_\bmq}^* Q_n G_d
+  \frac{1}{V} \sum_n P_n \hatG(\bmq_n) \nonumber \\
 &=&  \frac{1}{V} \sum_n P_n [\hatG(\bmq_n) -G_d] \nonumber \\
 &=&  \frac{1}{V} {\sum_{n \alpha}}^\prime ({\cal E}_{n \alpha}
 - {\cal D}_{n \alpha}).
\eeqa
The diagonal fluctuation is defined through,
\beq{c21}
  {\cal D}_{n \alpha}
    = \langle \chi_\alpha(\bmq_n) | G_d \rangle
    = \sum_j \chi_{\alpha,jj}(\bmq_n) f^0_j(1-f^0_j).
\eeq
Note the difference between (\ref{c11}) with $f=\rho/b$ and the present
equation where $f^0_i$ is the stationary solution of the generalized
Boltzmann equation (\ref{b8}).
In semi-detailed balance models $f^0_j=f$, but in multispeed LGA's lacking
semi-detailed balance $f^0_j$ differs from $f$ already in mean field
approximation (\ref{b7}).

Inspection of (\ref{c20}) shows that the $n \alpha$--summation
contains only as many
terms as there are global invariants; so $J$ in (\ref{c19}) is ${\cal
O}(V^{-1})$ and accounts for finite size effects. In the thermodynamic limit
$J$ vanishes and the ring equation (\ref{c19}) yields a closed
equation for the excess correlation function $C$. The diagonal part
$G_d$ then no longer appears in the ring equation.

\section*{Acknowledgements}
HJB is financially supported by the ``Stichting voor
Fundamenteel Onderzoek der Materie'' (FOM), which is sponsored
by the ``Nederlandse Organisatie voor Wetenschappelijk Onderzoek'' (NWO).
M.H.E. acknowledges  support from a Nato Travel Grant for visiting the
Physics Department of the University of Florida in the summer of 1993.
Similarly J.W.D. acknowledges FOM support for visiting the Institute of
Theoretical Physics at Utrecht University in the fall of 1993.


\newpage 

\end{document}